\documentclass{article}

\usepackage{arxiv}

\usepackage[utf8]{inputenc} % allow utf-8 input
\usepackage[T1]{fontenc}    % use 8-bit T1 fonts
\usepackage{url}            % simple URL typesetting
\usepackage{booktabs}       % professional-quality tables
\usepackage{amsfonts}       % blackboard math symbols
\usepackage{nicefrac}       % compact symbols for 1/2, etc.
\usepackage{microtype}      % microtypography
\usepackage{lipsum}
\usepackage{graphicx}
\usepackage[hidelinks]{hyperref}
\usepackage{adjustbox}
\usepackage{makecell}
\usepackage{subcaption,caption}

\newcommand{\revcutifc}[1]{\textcolor{red}{\sout{#1}}}

\renewcommand{\revcutifc}[1]{}

\title{An interpretable generative multimodal neuroimaging-genomics framework for decoding Alzheimer's disease}

\author{Giorgio Dolci$^{1,2,3}$, Federica Cruciani$^2$, Md Abdur Rahaman$^3$, Anees Abrol$^3$, Jiayu Chen$^3$, Zening Fu$^3$,\\ Ilaria Boscolo Galazzo$^2$, Gloria Menegaz$^{2,+}$, and Vince D. Calhoun$^{3,+}$,\\ for the Alzheimer’s Disease Neuroimaging Initiative$^*$ \\\\
$^1$Department of Computer Science, University of Verona, Verona, Italy\\
$^2$Department of Engineering for Innovation Medicine, University of Verona, Verona, Italy\\
$^3$Tri-Institutional Center for Translational Research in Neuroimaging and Data Science (TReNDS),\\ Georgia State University, Georgia Institute of Technology, Emory University, Atlanta, GA, USA\\\\
$^+$V.D. Calhoun and G. Menegaz equally contributed as last authors to this work.\\
$^*$Data used in preparation of this article were obtained from the Alzheimer’s Disease \\Neuroimaging Initiative (ADNI) database (adni.loni.usc.edu). As such, the investigators within the ADNI \\contributed to the design and implementation of ADNI and/or provided data but did not participate in analysis \\or writing of this report. A complete listing of ADNI investigators can be found at: \\\url{http://adni.loni.usc.edu/wp-content/uploads/how_to_apply/ADNI_Acknowledgement_List.pdf}}

\begin{document}
\maketitle
\begin{abstract}
\textbf{Objective:} Alzheimer's disease (AD) is the most prevalent form of dementia worldwide, encompassing a prodromal stage known as Mild Cognitive Impairment (MCI), where patients may either progress to AD or remain stable. The objective of the work was to capture structural and functional modulations of brain structure and function relying on multimodal MRI data and Single Nucleotide Polymorphisms, also in case of missing views, with the twofold goal of classifying AD patients versus healthy controls and detecting MCI converters.
\textbf{Approach:} We propose a multimodal DL-based classification framework where a generative module employing Cycle Generative Adversarial Networks was introduced in the latent space for imputing missing data (a common issue of multimodal approaches). Explainable AI method was then used to extract input features' relevance allowing for post-hoc validation and enhancing the interpretability of the learned representations.
\textbf{Main results:} Experimental results on two tasks, AD detection and MCI conversion, showed that our framework reached competitive performance in the state-of-the-art with an accuracy of $0.926\pm0.02$ and $0.711\pm0.01$ in the two tasks, respectively.
The interpretability analysis revealed gray matter modulations in cortical and subcortical brain areas typically associated with AD. Moreover, impairments in sensory-motor and visual resting state networks along the disease continuum, as well as genetic mutations defining biological processes linked to endocytosis, amyloid-beta, and cholesterol, were identified.
\textbf{Significance:} Our integrative and interpretable DL approach shows promising performance for AD detection and MCI prediction while shedding light on important biological insights.
\end{abstract}

% keywords can be removed
\keywords{Alzheimer's disease \and Generative model \and Imaging-genetics \and Explainable AI}

\section{Introduction}
\label{sec:introduction}

Alzheimer's disease (AD) is a chronic neurodegenerative disorder that affects millions of people worldwide (approximately 30 million in 2015 \cite{vos2016global}. It is the most common cause of cognitive impairment, gradually impacting the activities of a patient’s daily life. It is characterized by the progressive loss of cognitive and functional abilities, including memory, language, and executive functions \cite{robinson2020inclusive}, with a temporal progression. Amyloid accumulation represents the first event, followed by tau accumulation, hypometabolism (assessed with Positron Emission Tomography (PET)), atrophy, and cognitive decline \cite{breijyeh2020comprehensive,knopman2021alzheimer}.
It is hence evident that the pathology changes of AD actually begin several years before the first clinical symptoms. 
Therefore, a timely AD diagnosis is highly beneficial for optimizing patient care and enabling appropriate therapeutic interventions.
Mild cognitive impairment (MCI) is the intermediate stage from normal cognitive function to AD, hence representing an opportunity for an early targeting of the disease.
However, it includes a very heterogeneous class of patients, including subjects that will likely convert to AD, known as MCI converters (MCIc), with an estimated annual conversion rate around the $16.5\%$ \cite{petersen2010alzheimer}, and subjects that remain stable after several years, being part of the MCI non-converters (MCInc) group \cite{hamel2015trajectory}. 

Among the available neuroimaging technologies, structural Magnetic Resonance Imaging (sMRI) and resting-state functional MRI (rs-fMRI) have provided unprecedented opportunities for deriving biomarkers allowing the early diagnosis of AD. For instance, sMRI is currently a key part of the diagnostic criteria for the differential diagnosis and longitudinal monitoring of patients with dementia, enabling the estimation of brain atrophy. Several studies have consistently observed both global and local atrophic changes in AD, lying along the hippocampal pathway (entorhinal cortex, hippocampus, parahippocampal gyrus, and posterior cingulate cortex) in the early stages of the disease, while atrophy in temporal, parietal and frontal neocortices emerges at later stages being associated with neuronal loss as well as with language, visuospatial and behavioral impairments \cite{frisoni2010clinical,braskie2013recent}.
Rs-fMRI, in turn, indirectly measures neural activity by detecting changes in the Blood Oxygenation Level Dependent (BOLD) signals, which depend on the neurovascular coupling \cite{johnson2012brain}.
In particular, investigating functional connectivity (FC; inter-regional coupling), functional network connectivity (FNC; inter-network coupling), and functional networks from BOLD rs-fMRI provides a means for understanding the mechanisms and relevance of the functional relationships across brain regions. A growing body of rs-fMRI studies suggests that failure of specific resting-state networks (RSNs) is closely related to AD, with the default-mode network (DM) and the salience network (SN) playing a pivotal role and being disrupted before clinically evident symptoms \cite{pini2021breakdown}. Specific alterations in these functional networks have been reported in AD patients, with prominent FC decreased within the DM and increased FC in the SN \cite{zhou2010divergent, celone2006alterations}. Moreover, disconnections within and between the different RSNs have been consistently demonstrated, particularly over long connection distances \cite{liu2014impaired}. All of these factors have contributed to the widespread view of AD as a disconnection syndrome being characterized by a cascading network failure, beginning in the posterior DM and then shifting to other systems containing prominent connectivity hubs, possibly associated with amyloid accumulation \cite{jones2016cascading}. Moreover, increased evidence supports the view that tau depositions are also related to functional brain architecture and FC changes, supporting the view of transneuronal tau propagation in AD \cite{franzmeier2020functional}. 

AD also features a strong genetic component. Strategies to extract linked genotype traits are commonly based on Single Nucleotide Polymorphism (SNPs) analysis \cite{kwok1999single}, representing variations of single nucleotides in DNA sequences that vary from person to person \cite{kwok1999single} and are present in at least $1\%$ of the population. Several 
Genome-Wide Association Studies (GWAS) have identified more than 40 AD-associated genes/loci, which are likely to increase the risk of developing the disease \cite{wightman2021genome,lambert2013meta,kunkle2019genetic}. Among them, the apolipoprotein E (APOE), in particular the $\epsilon4$ allele, PICALM, CLU, ABCA7, and CR$1$ are the most important genes being associated with AD risk factor or the progression from MCI to AD \cite{winblad2004mild,tahami2022alzheimer,ungar2014apolipoprotein,xu2015role,jun2010meta,harold2009genome,corneveaux2010association, de2019role}.
 
This great heterogeneity of available biomarkers offers a unique opportunity to explore various aspects of the disease continuum. Each biomarker provides valuable information about specific characteristics, enabling a multifaceted investigation of AD which calls for methods that can effectively integrate and leverage complementary information. 
In this regard, artificial intelligence, Machine Learning (ML) and particularly Deep Learning (DL), emerge as promising technologies to tackle this complex task. 
Different ML algorithms have been developed for integrating heterogeneous data, such as Canonical Correlation Analysis (CCA), Partial Least Squares (PLS), and ensemble methods, e.g. consisting of different Support Vector Machines (SVM). Even if these methods can perform well in studying neurodegenerative diseases \cite{cruciani2024identifying, hao2017mining}, they have some limitations, such as requiring vectorized input data, hence denaturing the spatial information of 3D volumes (e.g., sMRI or PET 3D images), and potentially being too simple for solving complex tasks.
On the other hand, by employing multiple layers of processing, DL models allow extracting progressively higher-level and more informative features from input data, with the option of preserving spatial information, without the need of vectorizing the input features. Moreover, the inclusion of multiple data sources into a single model can uncover complex and deep non-linear associations across the input features from a multimodal perspective \cite{saxe2021if}.
As a result, multimodality is gaining significant popularity representing the key approach to derive valuable insights into complex and multifaceted neurodegenerative diseases such as AD \cite{abdelaziz2025multi,dolci2024multimodal,wu2023girus,rahaman2023deep,dolci2022deep,abrol2019multimodal}.
However, despite the promising foreseen of such an approach, multiple drawbacks are present and represent the focus of the current research. The main concern resides in missing data management. Particularly in the biomedical domain, it is very common to incur in missing values or acquisitions for certain subjects or entire study cohorts due to different reasons, such as missing acquisition, corrupted data, patients dropout from a study \cite{de2019deep}, and the requirements of privacy and expensive tests. Additionally, the interpretability of a model's predictions is a central characteristic of decision-aiding models, which is still not pervasively addressed. In the past years, many high-performing prediction models have been proposed lacking a clear rationale to be effectively considered for practical use. Strong and validated explanations associated with a given prediction are fundamental for increasing trust in the results as well as for their applicability in a real-world scenario. Innovative and viable missing modality management as well as interpretability are the key attributes that a multimodal model for diagnosis detection should have.

Diving into missing data management, the simplest and most common solutions consist either in discarding samples with missing modalities, or in filling the missing values with zeros \cite{venugopalan2021multimodal}, and computing imputations based on data interpolation \cite{ritter2015multimodal}. Such solutions are evidently suboptimal since they could significantly reduce the number of training samples, already small when addressing biomedical-related tasks, or introduce important biases in the data and the model due to the interpolation or the zero filling.
Alternative methods to exploit the information of all the available subjects were proposed. A possible approach is the complementation of incomplete data representation, which consists in extracting a latent representation from both the complete and incomplete modalities, avoiding the need for data imputation \cite{liu2021incomplete,zhong2019emerging}. Most recent approaches aim at generating missing data either in the input space \cite{gao2023multimodal,zhang2022bpgan,cai2018deep} or in an intermediate latent representation \cite{ye2022pairwise,gao2021task,zhou2019latent} relying on generative models such as Generative Adversarial Networks (GANs) \cite{tu2024multimodal,zhang2024pyramid,hwang2023real,gao2023multimodal,ye2022pairwise,gao2021task,goodfellow2020generative} and their variants, or Variational Autoencoders \cite{kingma2013auto}. This last approach has been successfully applied to the AD continuum investigation. It allows the exploitation of the availability of multimodal data for capturing the relationship among different data sources in the latent space, enabling the generation of one modality from another.
Generative models, hence, appear as the route to be followed when dealing with missing data. Interestingly, the recently proposed Cycle-consistent GANs (Cycle-GANs) \cite{zhu2017unpaired} have shown impressive results in various knowledge translation tasks since they offer a flexible and effective approach for learning mappings across different domains without relying on paired data. 

Moving to model interpretability, it is well established that with increasing model complexity, interpretability decreases drastically. However, in order to better understand the mechanisms that underlie the AD continuum and allow the models to be applied in clinical and real-life applications, it is vital to understand the reasons behind a certain output. In this case, eXplainable Artificial Intelligence (XAI), becomes essential to understand why a given model made a certain prediction. XAI encompasses \textit{post-hoc} methods that allow the assignment of an importance score to each feature, reflecting its role in the classification task. This means finally opening the `black box' of complex models hopefully increasing their exploitability in clinical practice. 
XAI is starting to be applied in multimodal frameworks to study neurodegenerative and psychiatric diseases. Few works could be found adopting simple gradient-based, feature perturbation methods, or attention weights \cite{dolci2024multimodal,zhang2024pyramid,gao2023multimodal,rahaman2021multi,el2021multilayer,hu2017analyzing}. On top of this, another overlooked yet stringent aspect is the strong validation of XAI methods and the obtained explanations. The evaluation of explanation methods is still under-investigated, however, since explainability is meant to increase confidence in AI, it is vital to systematically analyze the obtained results referring also to the prior knowledge derived from the state-of-the-art.

In this rapidly evolving landscape, where multimodality holds a central role, we aim at shading light on the importance of the complementary information that could be derived from advanced biomarkers such as brain FC and SNPs mutations, together with the well-established brain atrophy measures derived from sMRI, in detecting AD-related modulations. With this purpose, we will present a multimodal framework for AD detection and MCI conversion prediction, which addresses (i) heterogeneous data integration, (ii) missing data management, and (iii) interpretability. 
Regarding the former, our framework allows the integration of heterogeneous data featuring different dimensionality, e.g., 3D volumes and vectorized data, and nature, meaning from multi-domain, e.g., imaging and genetics.
Concerning the second, we propose an approach that allows generating missing modalities in the latent space obtained after input feature reduction, ensuring high accuracy in reconstructing latent features while minimizing computational demands. The goal is to define a framework that can be generalized to any missing modality, without requiring to have at least one specific modality shared by all the subjects in the considered population.
Finally, we aimed at emphasizing the interpretability of the proposed framework in order to reinforce its transparency and reliability by conducting an XAI post-hoc interpretability analysis, supplemented by a robust validation step, which enables to precisely discern the contribution of each input feature to the classification of subjects in the AD continuum, thus highlighting relevant phenotypic and genotypic biomarkers for AD.

\section{Materials and Methods}
\subsection{Dataset}
Data used in the preparation of this article were obtained from the Alzheimer’s Disease Neuroimaging Initiative (ADNI) database (\url{https://adni.loni.usc.edu/}). The ADNI was launched in 2003 as a public-private partnership, led by Principal Investigator Michael W. Weiner, MD. The primary goal of ADNI has been to test whether serial MRI, PET, other biological markers, and clinical and neuropsychological assessment can be combined to measure the progression of MCI and early AD. For up-to-date information, see \url{www.adni-info.org}.

%An accurate and extensive subject selection was performed to retain, for each subject, the baseline timepoint for all the selected imaging modalities, that is sMRI and rs-fMRI. 
For each subject, the baseline timepoint was retained for all the selected imaging modalities, that is, sMRI and rs-fMRI.
The respective genetic variants (SNPs) were then selected, if available. The final study cohort included $1911$ subjects, divided into healthy controls (CN), AD, MCI non-converter (MCInc), and MCI converter (MCIc). In detail, following the classification available on the ADNI website, an MCI subject was considered as MCIc if dementia was diagnosed at any timepoint after MCI diagnosis. Table \ref{table:demos} shows the Task 1 and Task 2 demographic information. For Task 1, the entire cohort, and the subset of individuals sharing both sMRI-fMRI and sMRI-SNPs are highlighted.

\begin{table*}[!ht]
    \centering
    \caption{Demographic information of the CN, AD, MCInc, and MCIc subjects. MCInc and MCIc have the following \% of missing SNPs 59.6\% and 30.8\%, respectively, and the missing fMRI are 60.1\% and 92\%, respectively. MMSE = Mini-Mental State Examination; wild (W) and mutated (M) alleles in APOE4: $0$ = homozygous (W/W), $1$ = heterozygous (W/M), $2$ = homozygous (M/M).}
    \begin{adjustbox}{width=\textwidth}
    \begin{tabular}{lccccccccccc}
        \toprule
        & \multicolumn{7}{c}{Task 1} & & \multicolumn{3}{c}{Task 2}\\
        & \multicolumn{3}{c}{CN} & & \multicolumn{3}{c}{AD} & & MCInc & & MCIc \\
        \cline{2-4} \cline{6-8} \cline{10-10} \cline{12-12} \\[-1.5ex]
        Cohorts & All subjects & sMRI-fMRI & sMRI-SNPs & & All subjects & sMRI-fMRI & sMRI-SNPs & & \multicolumn{3}{c}{Only for testing phase} \\
        \midrule
        Count & $644$ & $321$ & $253$ & & $332$ & $66$ & $152$ & & $646$ & & $289$ \\
        Age (y) & $73.6 \pm 6.6$ & $72.4 \pm 7.1$ & $76.4 \pm 5.4$ & & $75.1 \pm 7.9$ & $74.7 \pm 8.1$ & $75.8 \pm 7.7$ & & $73 \pm 8.5$ & & $74 \pm 8.5$ \\
        Gender (F) & 284 (360) & 117 (204) & 137 (116) & & 181 (151) & 35 (31) & 84 (68) & & 374 (272) & & 172 (117) \\
        MMSE & $29.1 \pm 1.1$ & $29.1 \pm 1.2$ & $29.1 \pm 1.1$ & & $23.1 \pm 2.2$ & $22.6 \pm 2.6$ & $23.3 \pm 2.0$ & & $27.9 \pm 1.8$ & & $26.9 \pm 1.8$ \\
        \makecell[l]{APOE e4\\($0/1/2$)} & $455/168/21$ & $224/87/10$ & $189/57/7$ & & $109/156/67$ & $19/34/13$ & $49/76/27$ & & $377/205/64$ & & $103/143/43$ \\
        \bottomrule
    \end{tabular}
    \label{table:demos}
    \end{adjustbox}
\end{table*}

3D T1-w MRI and rs-fMRI acquisitions were considered as imaging input channels and were acquired with the following sequence parameters: (i) sagittal accelerated MPRAGE, TR/TE = shortest, TI = $900$ ms, flip angle = $9^\circ$, Field Of View = $256\times256$ mm$^2$, spatial resolution = $1\times1\times1$ mm$^3$, slices = $176-211$), (ii) rs-fMRI: TR/TE =$ 3000/30$ ms, FA = $90^\circ$, FOV = $220\times220\times163$ mm$^3$, $3.4$-mm isotropic voxel size. $200$ fMRI volumes were acquired in almost all subjects, with minimal variations in a small subset (e.g., $197$ or $195$ volumes). More details about the data acquisition can be found in \cite{weiner2017alzheimer}. 
Concerning the genetic data, DNA samples were genotyped using Illumina Human610-Quad or Illumina HumanOmniExpress BeadChip.

\subsection{Preprocessing and feature engineering}
The sMRI volumes preprocessing included tissue segmentation in Gray Matter (GM), White matter, and Cerebrospinal fluid (CSF) using the modulated normalization algorithm. Only the GM volume was considered for this study, to which a smoothing using a Gaussian kernel (FWHM = $6$mm) was applied. The full preprocessed GM volume was used as input for the sMRI channel resulting in an input size of $121\times145\times121$ for each subject.
The rs-fMRI data was preprocessed using the statistical parametric mapping toolbox (SPM12, \url{http://www.fil.ion.ucl.ac.uk/spm/}) including rigid body motion correction to correct subject head motion, slice-timing correction, warping to the standard MNI space using the EPI template, resampling to ${(3mm)}^3$ isotropic voxels, and smoothing using a Gaussian kernel (FWHM = 6mm), following the preprocessing proposed in \cite{du2020neuromark}. Fifty-three independent components (ICs) covering the whole brain were extracted using spatially constrained ICA with the Neuromark\_fMRI\_$1.0$ template (available in the GIFT software; \url{http://trendscenter.org/software/gift}). For each subject, a correlation matrix was then created computing the Pearson correlation between the ICs time courses, resulting in a $53$x$53$ static functional network connectivity (sFNC) matrix. This was divided into $7$ RSNs, named: (i) Sub-cortical network (SC), (ii) Auditory network, (iii) Sensorimotor network (SM), (iv) Visual network (VI), (v) Cognitive-control network (CC), (vi) Default-mode network (DM), and (vii) Cerebellar network (CB) (please, refer to Supplementary Table 2 for more information). The upper triangular matrix was then vectorized, resulting in an input vector of 1378 features for each subject. Quality control of MRI images is detailed in the supplementary materials section "Materials and Methods", subsection "Preprocessing quality control of MRI".

Moving to the genomics data, pre-imputation QC was performed to remove Single Nucleotide Polymorphisms (SNPs) with minor allele frequency (MAF) $< 0.05$, call rate $< 0.98$, and Hardy Weinberg Equilibrium $< 10^{-3}$ (details in \url{https://www.synapse.org /#!Synapse:syn2290704/wiki/64710}). Imputation was performed with the same reference panel. Only the SNPs with imputation $r^2 > 0.4$ were included. Linkage disequilibrium (LD)-based pruning with $r^2 = 0.8$ in a window of $50$kb was applied, yielding $445838$ SNPs for further analyses \cite{jansen2019genome}. A feature selection leveraging on the genome-wide association study (GWAS) on AD conducted by \cite{wightman2021genome} was carried on to include all the relevant SNPs having GWAS \textit{p}-values less than $1$e$^{-04}$. A total of $565$ SNPs was selected and a value between 0 and 2 was assigned based on the presence of mutated alleles. In detail, defining wild and mutated alleles respectively as W and M, a score of $0$ means wild homozygous alleles (W/W), a score of $1$ indicates heterozygous alleles (W/M), and a score of $2$ defines mutated homozygous alleles (M/M).

\begin{figure*}[!ht]
    \centering
    \includegraphics[width=\textwidth]{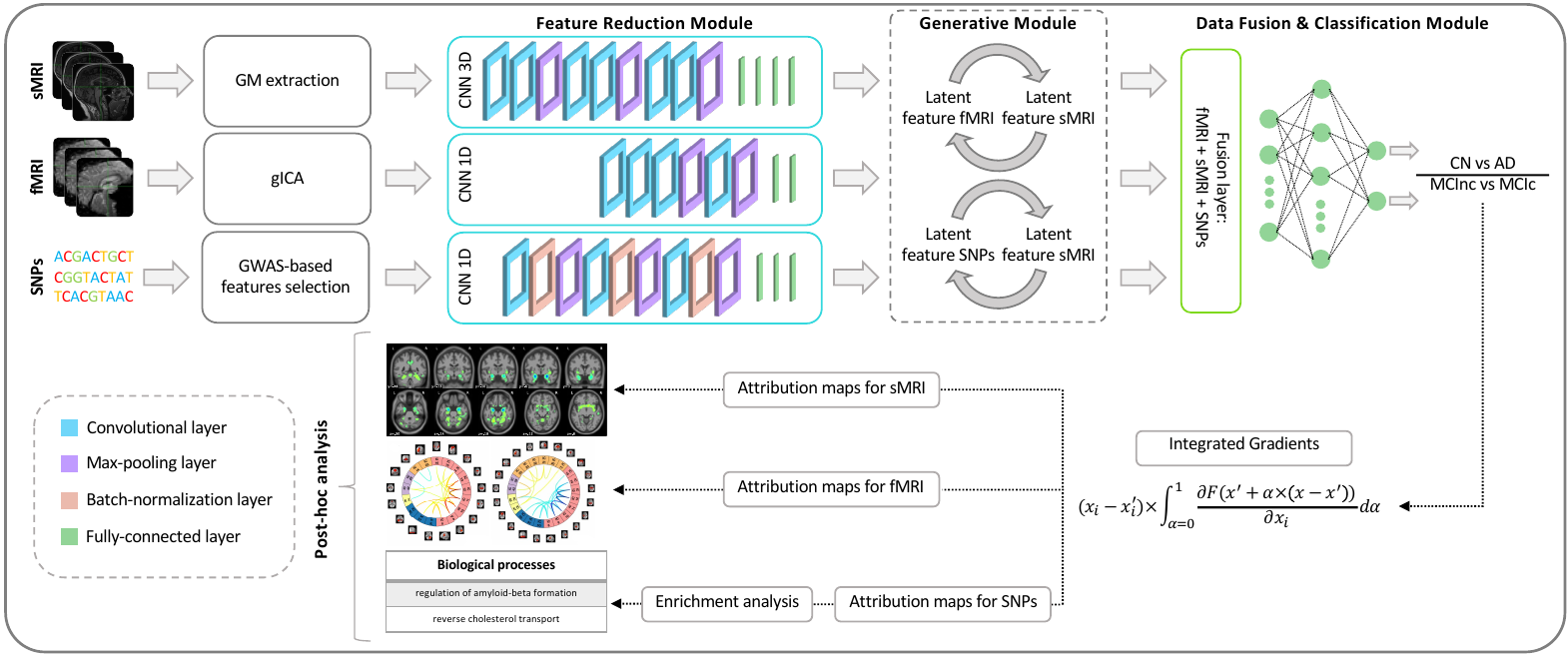}
    \caption{The proposed framework that integrates sMRI, fMRI, and SNPs is articulated in three modules: (i) A feature reduction module where the latent representations are extracted for the input data; (ii) A generative module where, if necessary, the missing modalities are imputed; and (iii) A data fusion \& classification module where the latent features are fused and then classified. Finally, \textit{post-hoc} XAI analysis is performed relying on the IG method for feature attribution derivation.}
    \label{fig:framework}
\end{figure*}

%\subsection{Classification tasks}
%\label{subssec:classification_tasks}
%Following the aim of this study, the proposed architecture was devised with a twofold classification aim: (1) AD detection, that is the differentiation of AD and CN subjects, also referred as Task 1 and (2) MCI conversion prediction, that is the stratification of MCIc and MCInc patients, also referred to as Task 2. In detail, the network was trained and tested on Task 1 and subsequently used to solve Task 2 allowing to assess its ability in discriminating different stages of disease and also in capturing and highlighting shared patterns across the different categories. 

\subsection{Framework architecture}
The proposed framework builds on our previous work \cite{dolci2022deep} and employed for addressing two classification tasks: Task 1, CN-AD, and Task 2, MCInc-MCIc, is shown in Figure \ref{fig:framework}. In detail, the architecture for disease detection consists of three modules: (i) \textit{Feature reduction module}, which performs a CNN-based feature extraction to derive a lower dimensionality latent space, separately for each input channel; (ii) \textit{Generative module} that, in the eventuality of missing modalities, actuates a generative process in the latent space transferring the knowledge from one domain to another; and (iii) \textit{Data fusion \& classification module} that fuses the latent features obtained for the three modalities and then performs the classification. Post-hoc interpretability analysis was then carried on in order to retrieve features' contributions to the classification task. The three modules will be detailed in the following paragraphs.

\paragraph{Feature reduction module}
The feature reduction module consists of three different CNNs, one for each input channel, leading to a latent low dimensionality representation consisting of $100$ latent features for each channel. 
The sMRI channel was analyzed through a 3D CNN defined by six convolutional layers (filter sizes of $3\times3\times3$ for the first four convolutional layers and $2\times2\times2$ for the last two, number of filters: $64$, $64$, $64$, $128$, $128$, and $128$) and three max-pooling layer, followed by four fully-connected layers (FCLs) (number of nodes: $1536$, $768$, $384$, and $192$).
The rs-fMRI channel was analyzed through a $1$D CNN, consisting of four convolutional layers (filter sizes $5$, number of filters: $64$, $128$, $128$, and $128$), two max-pooling layers, and two FCLs (number of nodes: $384$ and $100$).
Finally, for genetic data, a $1$D CNN was employed, enclosing three convolutional blocks, each including a sequence of convolutional layers (filter sizes $3$, number of filters: $64$, $64$, and $128$), batch-normalization and max-pooling, followed by three FCLs (number of nodes: $1024$, $512$, and $128$).
The ReLU activation function was used for all the layers of these three CNNs. 

\paragraph{Generative module}
The generative module allows imputing the missing modalities in the latent space given the others, when necessary. This task is possible thanks to the injection in the complete framework of four pre-trained generators derived from two different cGANs \cite{zhu2017unpaired}. The first, the sMRI-SNPs-cGAN allows imputing the latent genetics features transferring the knowledge from the latent sMRI one and \textit{vice-versa}, and the second, the sMRI-fMRI-cGAN, transfers the knowledge again from the latent sMRI features to generate the respective rs-fMRI ones, and \textit{vice-versa}. The scarcity of subjects sharing all modalities impeded the development of an fMRI-SNPs-cGAN model.

\begin{figure}[!ht]
    \centering
    \includegraphics[width=0.5\linewidth]{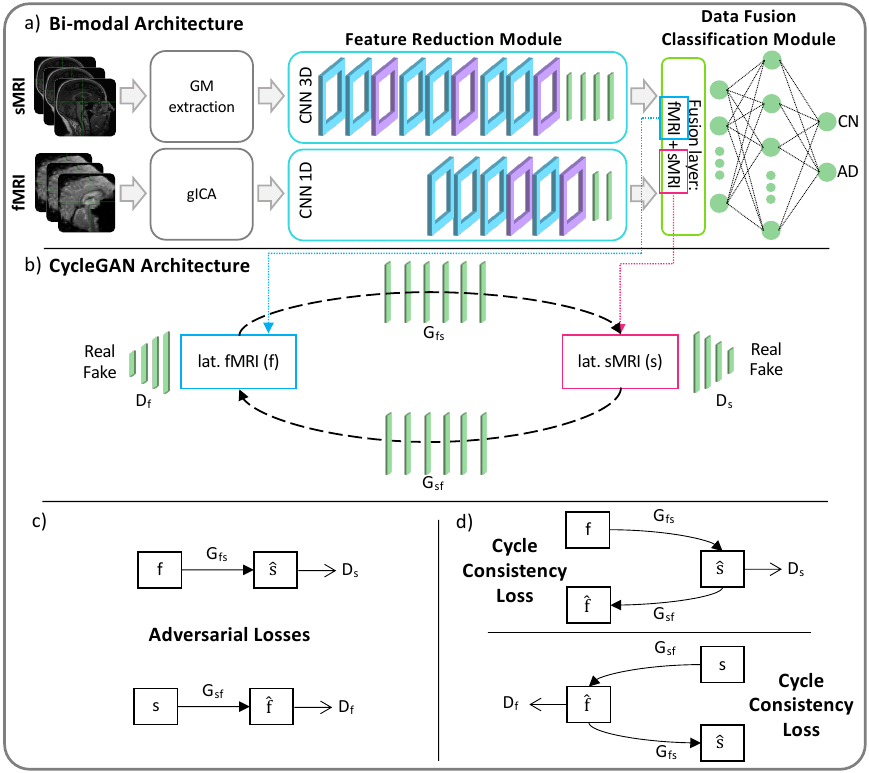}
    \caption{The bi-modal model sMRI-fMRI-NN/cGAN. a) The sMRI-fMRI-NN is composed of feature reduction and data fusion \& classification modules (equal architectures of the full model). b) After training, the feature reduction module extracts the latent features of both MRI used for training the sMRI-fMRI-cGAN, whose losses are described in c) and d).}
    \label{fig:bi-modal_framework}
\end{figure}

The cGANs were built and trained prior to the full framework. In detail, to obtain the pretrained generators, the first step required to obtain the channels' latent representations (i.e., the $100$ features vector for each modality) needed to subsequently train the cGANs. This was necessary since, as already highlighted, the proposed framework performs the data generation directly in the latent space and not in the original input space. %. 
Since the subset of subjects having all three modalities was not numerous enough to train a classification model, two separate bi-modal models were developed, one having sMRI and SNPs (sMRI-SNPs-NN) as input channels and the other having sMRI and rs-fMRI as input channels (sMRI-fMRI-NN). Figure \ref{fig:bi-modal_framework}a shows the schematic representation of the sMRI-fMRI-NN, as a reference example. %In order to obtain the same latent features as the complete framework, 
The feature reduction for each channel and the data fusion \& classification modules of these bi-modal models were the same as the ones adopted in the complete framework and described in the previous paragraph. 
Once the latent vectors were obtained, they were given as input to two cGANs, namely sMRI-fMRI-cGAN and sMRI-SNPs-cGAN, one for each bi-modal model, and hence for each information transfer, from sMRI to rs-fMRI and from sMRI to SNPs, respectively.
Figure \ref{fig:bi-modal_framework}b shows the sample sMRI-fMRI-cGAN architecture used to generate sMRI from rs-fMRI and \textit{vice-versa}. The generators are composed of six FCLs (number of nodes: $256$, $512$, $512$, $1024$, $512$, and $100$) activated by a ReLU activation function. The discriminators consisted of four FCLs (number of nodes: $256$, $128$, $64$, and $1$) activated by the LeakyRelu function, alternated with three dropout layers (dropout probability: $0.3$). The same architecture was considered for the sMRI-SNPs-cGAN.

\paragraph{Data fusion \& classification module} The data fusion \& classification module consists of a fusion layer and a classifier. The latent features obtained either from the data reduction module or the data generation module (for the subjects with missing modalities) are fused through vector concatenation, resulting in $300$ features. A multilayer perceptron composed of three FCLs (number of nodes: $150$, $75$, and $2$) activated by the ReLU function was then used for classification. Softmax activated the output layer.

\subsection{Training scheme}
\begin{figure}[!ht]
    \centering
    \includegraphics[width=0.5\linewidth]{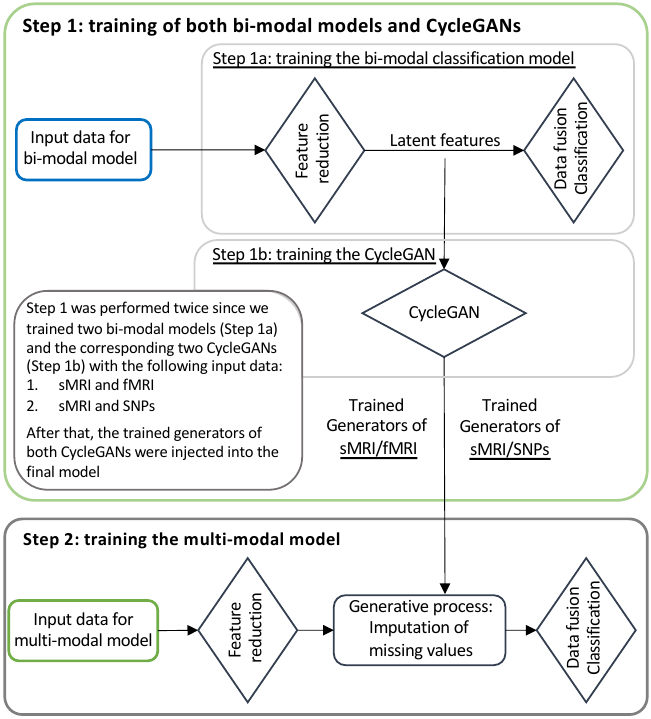}
    \caption{The framework training is performed in two steps: Step 1 involves the training of sMRI-fMRI-NN and sMRI-SNPs-NN (Step 1a) and the respective sMRI-fMRI-cGAN and sMRI-SNPs-cGAN (Step 1b); while in Step 2 the multimodal framework is trained.}
    \label{fig:training_workflow}
\end{figure}

Figure \ref{fig:training_workflow} illustrates the training workflow of the entire framework. As already specified, the training phase was not performed end-to-end but was divided into two steps, each one with end-to-end training, necessary for the correct training of the generation module: (i) Step 1: Training of bi-modal models and cGANs, and (ii) Step 2: Training of the full framework. 

Hyperparameter tuning was performed through a grid search. The number of layers, channels, nodes, and the filters' size were tuned considering the single-modality architectures for the classification (Task 1) before training the multimodal framework. 

\paragraph{Training Step 1: Bi-modal models and cGANs}\label{training_step_1}

In the training Step 1 the two sMRI-fMRI-NN and sMRI-SNPs-NN bi-modal models were trained for the AD versus CN classification (Step 1a, Figure \ref{fig:training_workflow}) in order to subsequently extract the latent features to be fed to the two cGANs, sMRI-fMRI-cGAN and sMRI-SNPs-cGAN (Step 1b, Figure \ref{fig:training_workflow}).
More in-depth, the bi-model models and subsequently the cGANs were trained relying on the AD and CN subjects sharing respectively the sMRI and rs-fMRI acquisition and the sMRI and SNPs acquisition, presented in Table \ref{table:demos}.
A $10$-folds stratified Cross Validation (CV) procedure for a total of $40$ epochs, Adam optimizer \cite{kingma2014adam} with a learning rate of $0.0001$, and mini-batch technique considering a batch size of $14$ and $9$ were selected, respectively, for the training of the sMRI-fMRI-NN and the sMRI-SNPs-NN. Weighted Cross Entropy (CE) was used as the loss function.
The sMRI-fMRI-NN and the sMRI-SNPs-NN achieved an average validation accuracy of $0.89\pm0.03$ and $0.864\pm0.03$, respectively.
The best model in terms of validation accuracy was retained for both the NNs and used to extract the $100$ latent features for each subject and for each modality. 
The newly obtained feature set was used to train the two cGANs. A mini-batch technique, keeping the same subjects in each batch as the bi-modal models, was used for both the sMRI-fMRI-cGAN and the sMRI-SNPs-cGAN training. 
Adam optimizer with a learning rate of $0.0001$ was used to train the generators for a total of $600$ epochs.
As in \cite{zhu2017unpaired}, the adversarial losses were considered as optimization targets for each cGAN as in Figure \ref{fig:bi-modal_framework}c. In detail, considering the sMRI-fMRI-cGAN as reference, given $s$ and $f$ the latent feature vectors for the sMRI and rs-fMRI, $G_{fs}$ and $G_{sf}$ the generators to obtain $s$ from $f$ and \textit{vice-versa} and $D_s$ and $D_f$ the discriminators to distinguish the $s$ from the generated $\hat{s}$ and the $f$ from the generated $\hat{f}$, respectively, the adversarial loss penalizes the $D_s$ and $D_f$ errors in discriminating between $\hat{s} = G_{fs}(f)$ and $s$, and $\hat{f} = G_{sf}(s)$ and $f$, respectively.
The major issue arising when considering this loss solely is that with a large enough capacity, the generators $G_{fs}$ and $G_{sf}$ can map the same set of input images to any random permutation of images in the target domain (latent features in our case) \cite{zhu2017unpaired}. The cycle consistency losses were hence introduced to limit the space of possible mapping functions by also penalizing the difference between $f$ and $G_{sf}(G_{fs}(f))$, and between $s$ and $G_{fs}(G_{sf}(s))$, respectively, hence completing the cycle (Figure \ref{fig:bi-modal_framework}d). The cycle consistency loss was used in both sMRI-fMRI-cGAN and sMRI-SNPs-cGAN in order to assess the performance of the generators in imputing missing modalities using the Mean Absolute Error (MAE).
The four pre-trained generators, two for the sMRI-fMRI-cGAN and two for the sMRI-SNPs-cGAN, were then extracted and injected in the latent space of the proposed multimodal framework for the online generation of the missing modalities during the full framework training.

\paragraph{Training Step 2: Full multimodal framework} 
The full multimodal framework including the three input channels, sMRI, rs-fMRI, and SNPs, was trained for classification Task 1, AD detection, relying on the complete AD and CN data cohort. 
Subjects were split into training and testing sets, respectively, including the $80\%$-$20\%$ of the study cohort. The testing set was kept unseen until the last testing phase.
A stratified $10$-fold CV procedure was applied to the training set. A mini-batch strategy ($12$ subjects per batch) was adopted during training. 
Adam optimizer with a learning rate of $0.0001$ was chosen and the model was trained for a total of $70$ epochs. Weighted CE was considered as the loss function. 
During this training phase, the four pre-trained cGAN generators' weights were kept frozen. Indeed, during backpropagation, only the weights of the feature reduction module and the data fusion and classification module were updated. Additional information about the training procedure is described in the supplementary materials "Materials and Methods" and subsection "Training scheme additional information".

\subsubsection{Testing and performance evaluation}
The best model, in terms of validation accuracy, obtained from the CV procedure of training Step 2 was considered for the testing phase for the two classification tasks. The full MCI cohort, never considered during the training phase, hence kept completely unseen by the model, was considered as a testing set for the classification Task 2, MCI conversion prediction.
The framework was finally evaluated in terms of accuracy (ACC), precision (PRE), and recall (REC) on both testing sets for the two classification tasks.
%The framework was finally evaluated in terms of accuracy (ACC), precision (PRE), recall (REC), F$1$ score, and Area Under the Curve (AUC) on both testing sets for the two classification tasks. %Area Under Precision Recall Curve (AUPRC),
%
Five independent runs reshuffling the train and test sets were performed for probing the generalization capabilities of the model, and the average test performance over the different runs was retained.

\subsection{Comparison with baseline approaches}
Two baseline methods with two different imputation strategies were implemented for the sake of comparison with the proposed framework in each Task. In detail, an ensemble SVM and an ensemble random forest classifiers with mean and zero imputation strategies were developed on the dataset considered in this study. Similar to the proposed framework, the cohort composed of CN and AD was used to train and test both ensemble models, and then Task 2 (MCI cohort) was employed only in the testing phase as an external test. Five independent runs, as in the proposed method, were performed to assess the ability to generalize on different testing splits.

\section{Post-hoc interpretability analysis}
The post-hoc interpretability analysis was performed relying on Integrated Gradients (IG) \cite{sundararajan2017axiomatic}, aiming at obtaining attribution maps describing the relevance of each input feature on the different input channels with respect to the outcome of the model. 
IG is a baseline attribution method, in the sense that, to obtain feature attributions, IG computes the path integral of the gradients along the straightline path from a given baseline $x'$ to the input $x$ \cite{sundararajan2017axiomatic} following the equation
\begin{equation}
    \centering
    IG_i(x)::=(x_i-x_{i}')\times\int_{\alpha=0}^{1} \frac{\partial F(x'_i+\alpha\times(x_i-x'_i))}{\partial x_i}d\alpha
\end{equation}
where $F$ is the model and $\alpha$ is a factor ranging in [$0,1$], determining the step size along the straightline path from $x$ to $x'$.
%
%This method satisfies three important axioms: (i) \textit{Sensitivity}, for which a null score is given to the input features which do not contribute to the prediction, (ii) \textit{Implementation invariance}, hence given a certain input, the derived attributions for two functionally equivalent networks are as well equivalent, and (iii) \textit{Completeness}, which ensures that the attributions approximately add up to the difference between the input and the baseline prediction scores \cite{sundararajan2017axiomatic}.
%
%IG attribution scores can assume both positive and negative values, where the larger the absolute value of the score of a given feature is, the higher its contribution to the prediction outcome is, either with a positive or negative effect.
%
%It is evident how the baseline choice impacts on the obtained attribution maps. Some recent studies have investigated this issue in order to raise awareness among the users and show the impact of different baselines on the IG scores obtained for the same input and model \cite{sturmfels2020visualizing, mamalakis2023carefully, dolci2023objective}. Of note, the main requirement for the baseline is to represent a neutral input for the model under investigation. Hence, it should produce a zero score or, equivalently, around $0.5$ prediction probability for both classes resulting from the activation function of the final classification layer. 
For the proposed model, a triplet of neutral baselines, one for each input data, was selected as follows: $3$D zero matrix, 1D zero vector, and 1D Gaussian noise vector for sMRI, rs-fMRI, and SNPs, respectively.

%\subsection{IG feature attribution analysis}
In our study IG was applied on the testing set of each classification task, resulting in a triplet of attribution maps for sMRI, rs-fMRI, and SNPs lying in the same space as the input data, for each subject and task.
The different maps will be referred to as sMRI-IG, fMRI-IG, and SNPs-IG in what follows. 

\subsubsection{Neuroimaging}
In order to retain the most relevant features, an initial sMRI-IG map thresholding was applied, selecting only the attribution values exceeding in absolute value the $99.5^{th}$ percentile of the respective IG values distribution voxelwise.
The average sMRI-IG values for $55$ cortical and subcortical brain regions based on the Harvard-Oxford atlas \cite{desikan2006automated} were then extracted for all the subjects in the testing sets of both classification tasks. Only the regions where at least $99\%$ of the subjects had average values exceeding the percentile threshold were kept for subsequent analysis.
For fMRI-IG, only the connections whose attribution scores were in absolute value above the $98^{th}$ percentile of the respective IG value distribution connection-wise were considered for further analysis. 
Different thresholds were chosen for the IG maps due to the different input sizes.

A statistical group-based analysis was then performed considering the attribution maps derived for each class, namely CN, MCInc, MCIc, and AD. Shapiro-Wilk test of normality was initially performed on the sMRI-IG and fMRI-IG derived features, revealing non-normally distributed values. The non-parametric Kruskal-Wallis test was hence performed to check the statistical difference across the four classes considered in each feature, derived either from rs-fMRI (most relevant connections resulting from fMRI-IG) or sMRI (most relevant brain regions derived from the sMRI-IG). When significance was found, the pairwise Wilcoxon Rank Sum Test was performed between all the possible pairs of the four considered groups (AD-CN, AD-MCIc, AD-MCInc, CN-MCIc, CN-MCInc, MCIc-MCInc) and adjusted for multiple comparisons with Bonferroni correction.
    
\subsubsection{Genetics}
Only the SNPs with attribution greater than the $65^{th}$ percentile in absolute terms of the SNPs-IG distribution, and those having positive attribution, were kept.
The Ensemble Variant Effect Predictor (VEP) tool \cite{mclaren2016ensembl} was used to annotate the selected SNPs in their corresponding genes, considering the default settings. 
%Selected settings included finding the co-located known variants and 1000 Genomes global minor allele frequency as frequency data for co-located variants parameters. No filtering was applied to the analysis to not exclude relevant biological data.
Enrichment analysis was then performed using the Gene Ontology enrichment analysis tool \cite{ashburner2000gene, gene2023gene, thomas2022panther} to derive the most important biological processes in which the annotated genes were involved. A background of $334$ genes was generated by annotating the $565$ SNPs-set used as input data for the model. Bonferroni correction for multiple testing was performed, and both corrected and no corrected \textit{p}-values were reported.
The SNPs annotation, as well as the enrichment analysis, was computed only on the classes of patients (AD, MCInc, and MCIc).

\section{Results}
%Classification performance and IG attribution analysis results will be firstly presented separately for classification Task 1 and Task 2. An overview of the overall group-based analysis will then follow.

\subsection{Task 1: AD detection}
The generators for both sMRI-fMRI-cGAN and sMRI-SNPs-cGAN models were tested by computing the MAE across the validation sets to compare the true latent space and the reconstructed one. For the sMRI-fMRI-cGAN, the generators achieved a MAE of $0.065\pm0.01$ and $0.074\pm0.01$ for the fMRI and the sMRI modalities, respectively. Meanwhile, the sMRI-SNPs-cGAN generators reached a MAE of $0.506\pm0.04$ and $0.333\pm0.10$ for SNPs and sMRI data, respectively.

The proposed framework reached $0.926\pm0.02$ in ACC, $0.910\pm0.05$ in PRE, and $0.876\pm0.03$ in REC for the differentiation between AD and CN subjects. Of note, the best model over the five generalization runs reached $0.964$ in ACC, $0.984$ in PRE, and $0.909$ in REC.
%The proposed framework reached $0.926\pm0.02$ in ACC, $0.910\pm0.05$ in PRE, $0.876\pm0.03$ in REC, $0.891\pm0.03$ in F$1$ score, and $0.960\pm0.01$ in AUC for the differentiation between AD and CN subjects. Of note, the best model over the five generalization runs reached $0.964$ in ACC, $0.984$ in PRE, $0.909$ in REC, $0.945$ in F$1$, and $0.970$ in AUC.%$0.829\pm0.03$ in AUPRC, $0.876$ in AUPRC,

Regarding the comparison with the baseline approaches for Task 1, the SVM with mean and zero imputation strategy achieved average ACCs of $0.699\pm0.01$ and $0.689\pm0.02$, respectively, while the random forest methods with still mean and zero missing data management approaches obtained average ACCs of $0.663\pm0.01$ and $0.663\pm0.01$, respectively, hence highlighting how our proposed framework outperformed simplest approaches. More information about the comparison with the baseline methods is reported in the supplementary materials "Results" section, "Classification comparison with baseline approaches" subsection.

\subsubsection{Feature relevance}
\begin{figure*}[!ht]
    \centering
    \includegraphics[width=\textwidth]{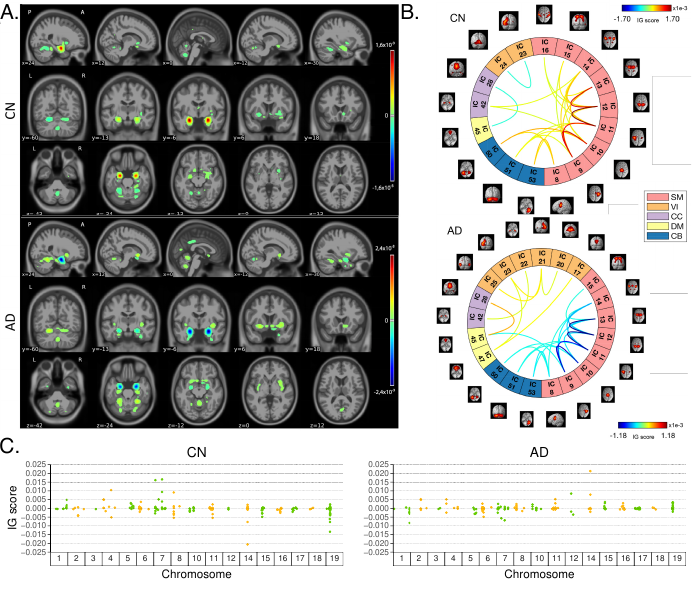}
    \caption{IG attribution maps for the classification Task 1. All the IG maps are presented for the correctly classified CN and AD subjects in the testing set. A. Average sMRI-IG maps, thresholded to retain IG scores exceeding the $99.5^{th}$ percentile; B. Average fMRI-IG derived connectograms, keeping only the connections over the $98^{th}$ percentile; C. Average SNPs-IG scores highlighting the SNPs with an associated IG score exceeding the $65^{th}$ percentile in absolute terms and grouped by chromosome.}
    \label{fig:IG_maps_task1}
\end{figure*}

Figure \ref{fig:IG_maps_task1} shows the average IG maps obtained for the AD detection task and for each input channel, averaged over the correctly classified subjects per class (test set). The \textit{jet} colorbar was used to highlight the attribution scores, where red and blue correspond to positive and negative attribution values, respectively.
In detail, Figure \ref{fig:IG_maps_task1}A shows the average sMRI-IG maps for the correctly classified CN and AD patients, overlaid onto the MNI$152$ template (1.5 mm). 
%A thresholding was applied in order to visualize only the relevance scores above the $99.5^{th}$ percentile of the relevance distribution, followed by a Gaussian smoothing with FWHM = 3 mm for visualization purposes. 
It is evident that the subcortical regions, in particular the hippocampus, are associated with high IG scores in absolute value, with the highest positive (negative) values for CN (AD). Cortical regions generally showed almost null relevance for both the CN and the AD-derived sMRI-IG maps. Of note, higher absolute values were found for the AD maps compared with the CN ones.

Moving to the fMRI-IG qualitative analysis, Figure \ref{fig:IG_maps_task1}B shows the average connectograms for CN and AD subjects, including only the connections above the $98^{th}$ percentile of the relevance distribution. The ICs-brain region correspondences are detailed in supplementary Table 3. 
The most relevant RSNs were the primary information processing-related networks (SM and VI) followed by multisensory integration networks (CC and DM), and cerebellum (CB), showing high relevance for both the CN and the AD-derived fMRI-IG. 
In particular, the $53$\% of the most relevant connections for the CN group belong to the SM network, while a percentage of $38$\% was found for the AD-derived fMRI-IG maps. On the other hand, the VI was highly involved for AD subjects, with the $28$\% of relevant connections belonging to this RSN, differently from the CN where only $2$ VI ICs resulted as relevant. A total of $15$ ICs were marked as relevant for both the CN and the AD-derived fMRI-IG, but with an opposite sign.
More in detail of the relevant connections between the different ICs, the CN subjects showed positive relevant intra-network connections in the CB and SM RSNs, with particularly high scores for the connections involving the post/paracentral and parietal gyri (ICs $9,10,11,12,13$). Inter-network positively relevant connections were also found between the SM and the CB, and the SM and the CC RSNs. Only two negative IG scores were instead retrieved, related to an intra-network connection in the VI network and to an inter-network connection between CC and DM. 
A similar pattern was found for the AD fMRI-IG relevant connections but with generally opposite IG-associated scores. In detail, negative relevance was mainly found for both the intra- and inter-network connections encompassing the SM and CB RSNs, with an intra-connection in SM (ICs 11-12) showing the highest negative relevance between the same ICs highlighted with the opposite sign in the CN. On the contrary, positive relevance was recorded for the connections between multiple VI ICs, in particular involving right middle occipital gyrus (IC $21$) with cuneus (IC $20$), inferior occipital gyrus (IC $23$), and middle temporal gyrus (IC $25$) as well as the inter-network connections VI-DM and CC-DM, with the latter being the most relevant one. 

Finally, concerning the SNPs-IG qualitative analysis, Figure \ref{fig:IG_maps_task1}C presents a Manhattan plot including the SNPs with an IG score higher than the $65^{th}$ percentile in absolute terms, grouped by chromosome. The y-axis reports the associated IG score. 
A complementary trend between the IG values associated with the SNPs was found between the two classes. This was particularly evident for Chr $1$, $7$, $11$, $14$, $15$, and $19$. A generally high involvement of the Chr $7$, $8$, $11$, $14$, and $19$ was present for both the AD and CN, with the majority of SNPs selected in these Chr showing high IG scores, either positively or negatively. Of interest, the most relevant SNPs, with positive IG were found in Chr $4$, $7$, and $8$ for the CN and in Chr $12$ and $14$ for the AD. On the contrary, the most negatively relevant SNPs for CN were found in Chr $14$ and $19$, as opposed to AD where they were mainly in Chr $1$, $6$, and $7$. 

\subsection{Task 2: MCI conversion prediction}
The MCI prediction task was performed by testing the full MCI cohort on the best model obtained for the AD detection task, for each one of the five independent runs. This allowed to assess its viability in predicting MCI conversion to AD, by correctly stratifying MCIc and MCInc subjects, without being trained for the specific task. 
This procedure allowed to reach an ACC of $0.711\pm0.01$, a PRE of $0.558\pm0.03$, and a REC of $0.610\pm0.03$ for the differentiation between MCIc and MCInc. Regarding the performance obtained using the best model trained on Task 1, we achieved $0.711$ in ACC, $0.612$ in PRE, and $0.550$ in REC.
%This procedure allowed to reach an ACC of $0.711\pm0.01$, a PRE of $0.558\pm0.03$, a REC of $0.610\pm0.03$, $0.581\pm0.01$ in F$1$ score, and finally an AUC of $0.755\pm0.01$ for the differentiation between MCIc and MCInc. Regarding the performance obtained using the best model trained on the classification Task 1, we achieved $0.711$ in ACC, $0.612$ in PRE, $0.550$ in REC, $0.579$ in F$1$ score, and $0.766$ in AUC. %$0.470\pm0.02$ in AUPRC, $0.505$ in AUPRC,

The baseline approaches were tested under the same training strategy as the proposed model also for Task 2, where the SVM with mean and zero imputation strategy obtained ACCs of $0.690\pm0.01$ and $0.688\pm0.01$, respectively, and ACCs of $0.698\pm0.01$ and $0.694\pm0.01$ for random forest with mean and zero strategies, respectively. Additional details related to the comparison are reported in supplementary materials "Results" section, "Classification comparison with baseline approaches" subsection.

\subsubsection{Feature relevance}
\begin{figure*}[!ht]
    \centering
    \includegraphics[width=\textwidth]{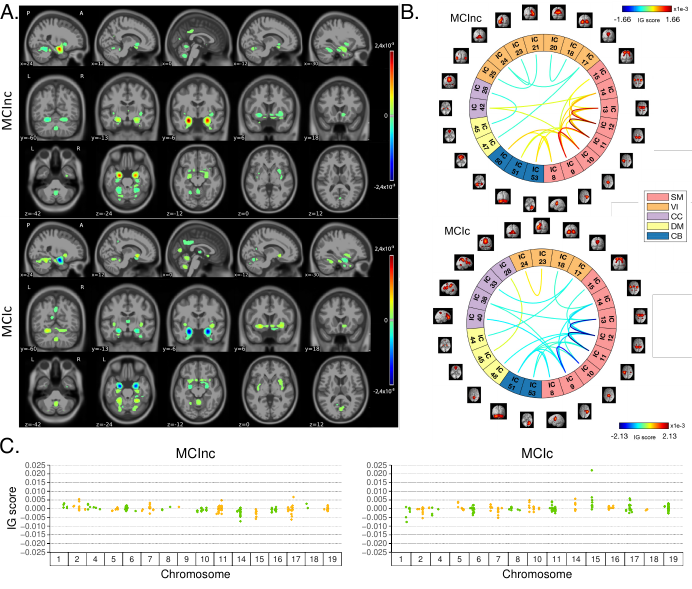}
    \caption{IG attribution maps for the classification Task 2. All the IG maps are presented for the correctly classified MCIc and MCInc subjects in the testing set. A. Average sMRI-IG maps, thresholded to retain IG scores exceeding the $99.5^{th}$ percentile; B. Average fMRI-IG derived connectograms, keeping only the connection over the $98^{th}$ percentile; C. Average SNPs-IG scores highlighting the SNPs with an associated IG score exceeding the $65^{th}$ percentile in absolute terms and grouped by chromosome.}
    \label{fig:IG_maps_task2}
\end{figure*}
In parallel with the results shown for Task 1, Figure \ref{fig:IG_maps_task2} shows a summary of the IG results obtained for the MCI conversion prediction task. The same visualization techniques presented for Task 1 were applied also for this figure. The maps were averaged over the correctly classified subjects for each class, and thresholding was applied to retain only the values above the $99.5^{th}$, $98^{th}$, and $65^{th}$ percentile of the relevance distribution, respectively for the sMRI-IG, fMRI-IG, and the SNPs-IG. 
%The sMRI-IG maps were smoothed and overlaid onto the MNI$152$ template. 
In detail, Figure \ref{fig:IG_maps_task2}A shows the average sMRI-IG maps for the correctly classified MCIc and MCInc patients. Similarly to what was shown for Task 1, the subcortical regions held the highest relevance, with the hippocampus being clearly highlighted as the most important region with positive attributions for MCInc and negative values for MCIc. Cortical regions did not show particular relevance, except for some temporal and occipital areas, which however showed low IG scores.

Moving to the fMRI-IG, Figure \ref{fig:IG_maps_task2}B shows the averaged connectograms. 
In accordance with Task 1, the same five RSNs resulted as the most relevant also for the MCI prediction conversion task, namely the SM, VI, CC, DM and CB. 
The MCInc fMRI-IG showed more involvement of VI ($25$\% of the relevant ICs), compared with the MCIc one ($14$\% of the relevant ICs). On the contrary, the MCIc IG connectome showed a higher number of ICs belonging to the CC network ($4$ ICs) compared to the two found for the MCInc one. 
The two connectograms had $11$ common connections with opposite trends.
More in details of the relevant connections between the different ICs, of interest, the most relevant connections between the ICs in the SM were exactly the same for MCIc and MCInc, with opposite signs, and mainly involved the post/paracentral and parietal gyri (ICs $9,10,11,12,13$), as for Task 1.
Moreover, the fMRI-IG for the MCInc subjects showed positive relevance also for the intra-network connections in the CB, and for the inter-network connections involving CB-SM and CC-SM. On the other hand, negative relevance was found for the intra-network connections in the VI, mainly involving occipital, lingual, calcarine, middle temporal gyri, and cuneus areas (ICs 17-24, 18-20, 21-25, 23-24), and for the inter-network connections between VI and DM, and CC and DM. 
A slightly different pattern was found for the MCIc fMRI-IG, whose relevant connections generally had a negative IG-associated score. In detail, intra-network connections with negative scores were found for SM and CB, as already presented, while inter-network negative connections were depicted for VI-CC, VI-DM, and SM-DM. Of interest, negative connections between CC and SM were also retrieved, never reported for the other classes in the study. Positive attribution was found only for an intra-network connection in the VI (ICs 23-24) and a connection between CC and DM (ICs 28-45). 

Finally, concerning the SNPs-IG, Figure \ref{fig:IG_maps_task2}C presents a Manhattan plot highlighting the SNPs with an IG score higher than the $65^{th}$ percentile in absolute terms. 
As expected and as already seen for the other modalities, a complementary trend between the IG values associated with the SNPs was found for the two classes, with positive weights being associated with the MCInc and negative weights associated with the MCIc for the same SNPs and \textit{vice-versa}. This was particularly evident for Chr $1$, $2$, $7$, $14$, $15$ and $17$. A generally high involvement of the Chr $11$ and $17$ was present for both the MCIc and MCInc, with the majority of SNPs showing high IG scores. The most relevant SNPs, with positive IG were found in Chr $17$ for the MCInc and in Chr $15$ for the MCIc. On the contrary, the most negatively relevant SNPs for MCInc were found in Chr $14$, $15$, and $17$, as opposed to the MCIc where they were mainly in Chr $1$, $2$, and $7$. 

\subsection{Group-based statistical IG analysis}
%The IG scores obtained for the four groups of subjects in the study were analyzed in order to quantify the differences and similarities outlined in the qualitative analysis. In what follows, the statistical analyses will be presented separately for the neuroimaging (sMRI and rs-fMRI) and the genetic channels.    

\paragraph{Neuroimaging modalities}
\begin{figure*}[!ht]
    \centering
    \includegraphics[width=\textwidth]{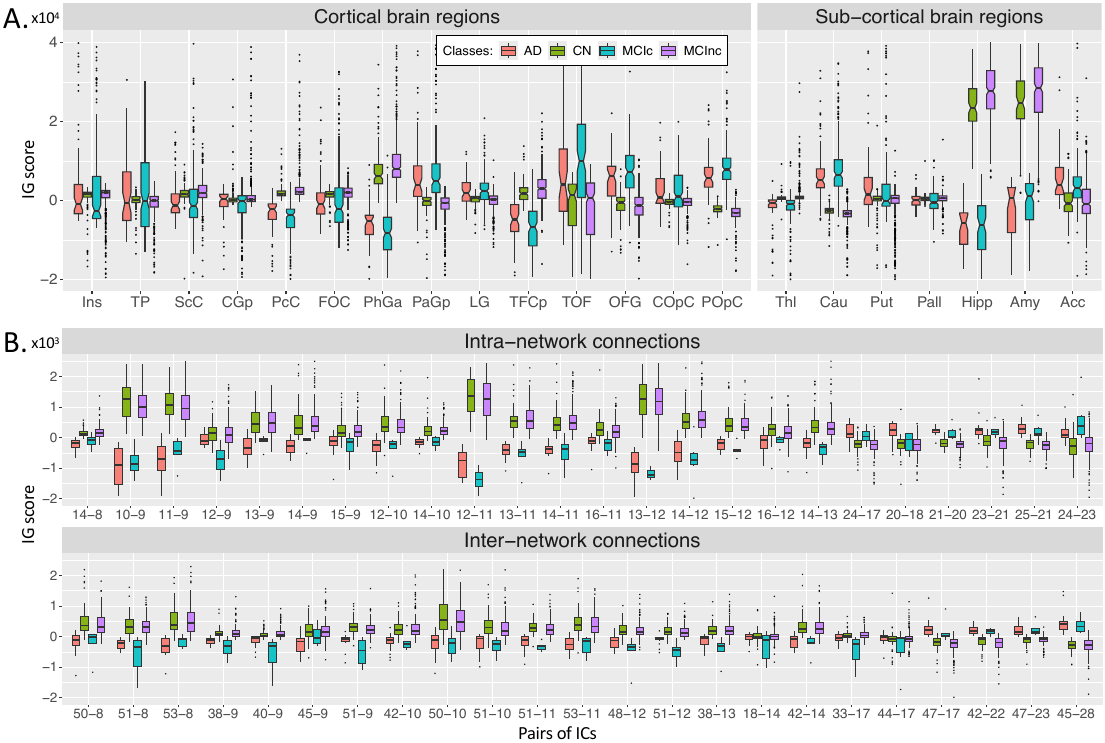}
    \caption{Boxplots showing the CN, AD, MCInc, and MCIc correctly classified subjects distribution in A. the most relevant cortical and subcortical brain regions resulted from the sMRI-IG, and B. the most relevant pairs of ICs for both intra- and inter-network connections.}
    \label{fig:srmi_fmri_boxplots}
\end{figure*}
\begin{figure*}[!ht]
    \centering
    \includegraphics[width=\textwidth]{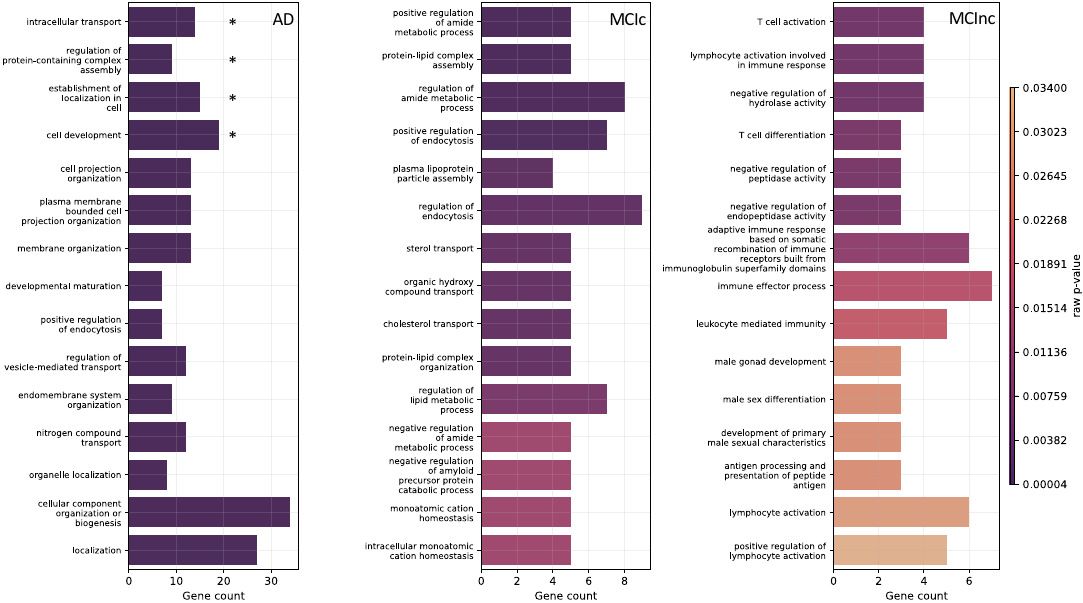}
    \caption{The top $15$ biological processes from enrichment analysis reported for the AD, MCIc, and MCInc groups (columns). The colorbar represents the \textit{p}-value, dark violet the lowest, while the number of genes involved is reported as the bar height. $*$ indicates biological processes surviving Bonferroni correction.}
    \label{fig:GEN}
\end{figure*}
Figure \ref{fig:srmi_fmri_boxplots}A shows the boxplots representing the distribution of the average sMRI-IG values in the most relevant cortical and subcortical brain regions (ROIs), and only those regions where at least the $99\%$ of the subjects had an IG value were retained. The complete acronym list is reported in supplementary Table 2.
A general agreement between the CN and the MCInc subjects, as well as between the AD and the MCIc patients is evident in all the considered ROIs, with the AD/MCIc subjects showing a generally higher variance. As expected from the qualitative analysis, the Hipp and the Amy resulted as the highest relevant ROIs, as well as the ones showing the highest group distance, with high positive attributions for the CN and the MCInc, and strong negative attributions for the AD and MCIc. Among the other subcortical ROIs, the Acc showed a notable relevance for all the groups, with the MCInc and CN having negative scores and the MCIc and AD positive ones. Cau and Put had an associated high positive relevance score for AD and MCIc, while almost null scores for the other two groups. Interestingly, among the cortical regions, the TOF was the most relevant for both the MCIc and AD (positive IG scores) and the MCInc and CN (negative IG scores). The largest distance between the group IG score distributions was found for PhGa, followed by the TFCp with the AD and MCIc having high negative values. On the contrary, the OFG, the POpC, and the PaGp showed high positive relevance for AD and MCIc patients associated with negative scores for the CN and MCInc.
The Kruskal-Wallis test was performed in order to establish whether differences, separately for each ROI, were significant considering the four groups together. The significant ROIs (all except the Ins and the TP) were further analyzed to investigate the group-related differences through the Wilcoxon Rank Sum test. The most significant differences were found between MCIc and MCInc in the subcortical ROIs, with the lowest \textit{p}-values being recorded for Hipp and Amy. Moreover, significant differences were also recorded for the temporal (PcC, PhGa, PaGp, TFCp) and occipital (TOF, OFG, POpC) cortical regions, with the most significant difference being recorded for PcC and PhGa. A coherent significance pattern was found between the contrasts MCIc-CN, MCInc-AD, and the CN-AD with generally higher \textit{p}-values. Of interest, no statistically significant differences were found in the sMRI-IG score distributions between AD and MCIc patients, except for PcC, TFCp, and POpC, while some significant differences were found between CN and MCInc but with relatively high \textit{p}-values. Supplementary Figure 1A shows the \textit{p}-values of statistical analysis between groups for each region.

Figure \ref{fig:srmi_fmri_boxplots}B shows the boxplots of fMRI-IG values for the intra- and inter-connections of the four groups. The same pattern of sMRI regarding the similarity between CN and MCInc, and AD and MCIc has been also highlighted for the rs-fMRI intra- and inter-network connections. Among the intra-network connections, the ones related to the sensorimotor area were the most important, such as IC9-IC10, IC9-IC11, IC11-IC12, and IC12-IC13. Regarding the inter-network connections, they showed generally lower relevance with respect to the intra-network ones. The same statistical analysis was carried out for fMRI-IG, considering all the most relevant connections resulting from the fMRI-IG scores for all the considered classes. The connections between IC$18$ (VI) – IC$14$ (SM) and IC$44$ (DM) – IC$17$ (VI) did not reach significance at the Kruskal-Wallis test, therefore they were excluded from the post-hoc analysis.
Differently from the sMRI-IG significance pattern, the most significant differences were recorded between the CN-AD and MCInc-AD contrasts, which showed significant differences for almost all the considered connections. In particular, for the MCInc-AD contrast, significant intra-network connections were part of the SM, involving connections in the parietal lobe, and in the VI RSNs. Concerning the inter-network connection, significance was found between ICs in the DM, part of the cingulate gyrus (IC47, IC45), and ICs in the CC or VI RSNs, as well as between ICs in the SM, located in the parietal lobe and ICs in the CC or CB RSNs. MCIc-CN and MCInc-MCIc showed the same significant pattern as the CN-AD contrast, showing generally higher \textit{p}-values. Of interest, as for the sMRI-IG statistical analysis, no significant differences were detected for the contrasts MCIc-AD and MCInc-CN, except for two connections between CB and SM (IC51-IC9 and the IC51-IC12 respectively for the MCInc-CN and the MCIc-AD). 
Supplementary Table 3 lists the ICs and corresponding brain regions.
The Bonferroni-corrected pairwise tests (Wilcoxon rank sum) are reported in supplementary Figure 1B, separately for intra- and inter-network connections, with the \textit{p}-values being represented in logarithmic scale.

\paragraph{Genetic modality} 
Figure \ref{fig:GEN} shows the biological processes derived from enrichment analysis for AD, MCIc, and MCInc classes as well as their corresponding raw \textit{p}-values. Only the SNPs having positive attributions higher than the $65^{th}$ percentile were retained for gene annotation. 
The left plot of Figure \ref{fig:GEN}, regarding AD patients, highlights the four AD-related biological processes that were statistically significant after \textit{p}-value correction, that is: \textit{intracellular transport} ($p_{\mathrm{bonf}}=0.0225$), regulation of protein-containing complex assembly ($p_{\mathrm{bonf}}=0.0413$), establishment of localization in cell ($p_{\mathrm{bonf}}=0.0446$), and cell development ($p_{\mathrm{bonf}}=0.0470$).
The other AD-related biological processes that emerged involved the endocytosis mechanism, hence processes related to the transport, such as \textit{positive regulation of endocytosis}, \textit{regulation of vesicle-mediated transport}, and \textit{nitrogen compound transport}. Additionally, biological processes related to the membrane and cell organization were detected.
%, such as \textit{cell organization}, \textit{cell projection organization}, \textit{cellular component organization or biogenesis}, \textit{organelle organization}, \textit{membrane organization}, \textit{plasma membrane bounded cell projection organization}, and \textit{endomembrane system organization}.
Similarly, the central plot, concerning the MCIc-related biological processes, also identifies the endocytosis mechanism, such as \textit{regulation of endocytosis} and \textit{positive regulation of endocytosis}. Of note, one process was related to the amyloid-$\beta$ (A$\beta$), that was \textit{negative regulation of amyloid precursor protein catabolic process}. Furthermore, important biological processes related to cholesterol, like \textit{sterol transport} and \textit{cholesterol transport}, and processes involving lipids, amide compound, and proteins,
%such as \textit{protein-lipid complex assembly}, \textit{plasma lipoprotein particle assembly}, \textit{protein-lipid complex organization}, and \textit{regulation of lipid metabolic process} 
were also highlighted. 
%Different processes involved the amide compound, e.i., \textit{positive regulation of amide metabolic process}, \textit{regulation of amide metabolic process}, and \textit{negative regulation of amide metabolic process}. 
The ten most frequent genes in terms of occurrences in the enriched biological processes for AD and MCIc patients were CLU, APOE, PICALM, APOA2, ABCA7, TREM106B, NECTIN2, TREM2, BIN1, and BLOC1S3. 
Regarding the MCInc, right plot, two biological processes were related to the T cell, the \textit{T cell activation} and the \textit{T cell differentiation}, while different processes involved the white blood cells and immunity. 
%such as \textit{lymphocyte activation involved in immune response}, \textit{lymphocyte activation}, \textit{positive regulation of lymphocyte activation}, \textit{adaptive immune response based on somatic recombination of immune receptors built from immunoglobulin superfamily domains}, \textit{immune effector process}, and \textit{leukocyte mediated immunity}. 
For this group, the five most frequent genes were FCER1G, ERCC1, CR1, RELB, and ACE.
Supplementary Tables 4, 5, and 6 provide additional information, such as the most frequent genes, the GO index, and the corresponding \textit{p}-value for each highlighted biological process for AD, MCIc, and MCInc, respectively.

\section{Discussion}
In this work, we proposed a multimodal generative and interpretable method, which holds the potential of addressing missing data management while focusing on model interpretability. We specifically applied this method to the crucial tasks of segregating AD patients from CN, and assessed its viability in detecting MCI conversion to AD by integrating neuroimaging (sMRI and rs-fMRI) and genetics (SNPs) data.
%through DL-based feature extraction, we introduced a multimodal data fusion framework, relying on input data not yet simultaneously investigated in the AD continuum in the current literature. 
Incomplete data is handled by generating missing modalities in the latent space, obtained after feature reduction. This ensures high accuracy in reconstructing latent features while minimizing computational demands. %To accomplish this, we relied on pre-trained generators from two cGANs, one trained to generate sMRI and SNPs data, and the other to generate sMRI and rs-fMRI patterns. 
However, the proposed approach is generalizable to any missing modality, providing a viable and effective solution for overcoming the missing data bottleneck. 
%The framework is trained to detect AD in a cohort including CN and AD subjects, while its generalization capabilities were then tested by predicting MCI conversion to AD in an independent study cohort. 
Furthermore, we conducted a post-hoc interpretability analysis, supplemented by a robust validation step, consolidating the findings' impact in the field.
%The next Subsections are dedicated, respectively, to the assessment of the classification performance and missing data processing (Subsection \ref{SS1}),  and to interpretability analysis in the light of neurosciences (Subsection \ref{SS2}).
    
\subsection{Classification performance and missing data management}
\label{SS1}
Focusing on Task 1, AD detection, our model reached competitive performance in the state-of-the-art, obtaining an average accuracy of $0.926\pm0.02$ on the testing set, with the best model reaching an accuracy score of $0.964$. Noteworthy, this result was achieved on a cohort where only the $6.5\%$ of the subjects had complete data across all three modalities. 
The proposed framework in the same Task 1 also outperformed the two baseline methods with the two imputation strategies, where they achieved an average accuracy of $0.699\pm0.01$ and $0.663\pm0.01$ for the ensemble SVM and ensemble random forest with mean imputation strategy, respectively, while accuracies of $0.689\pm0.02$ and $0.663\pm0.01$ were obtained using the zero imputation strategy for SVM and random forest, respectively.
Our dataset was constructed considering the subjects having at least the sMRI modality, however, the proposed framework showcases the capability to potentially impute all three modalities by transferring knowledge across domains, not being limited to having at least one acquisition for all the subjects. Indeed, the injection of the pre-trained generators, that is the two cGANs, allows both to impute the missing rs-fMRI or SNPs from the sMRI and the \textit{vice-versa}, meaning that it would allow deriving the missing sMRI from the other two modalities, if needed. This allows relaxing the requirement of having at least one acquisition for all the subjects which is one of the strengths of the proposed model compared with the state-of-the-art.

The simplest model, Venugopalan et al. \cite{venugopalan2021multimodal} proposed a DL-based multimodal method based on sMRI, clinical, and genetics data, excluding features if missing in more than $70\%$ of subjects, and filling the remaining missing data with zeros. Their classification task was different from ours since they included the MCI subjects in a three-class classification problem. While they achieved an accuracy of $0.780$ for the classification between CN, MCI, and AD, the accuracy dropped to $0.630$ when classifying CN from the full patient cohort using only sMRI and genetics. Moreover, the proposed technique had some drawbacks, mainly linked to the biases introduced in the network, as already discussed.
Alternatively, methods to exploit the information of all the available subjects, without explicitly generating the missing modalities, were proposed relying on latent representation learning. 
Liu et al. \cite{liu2021incomplete} developed an Auto-Encoder-based multi-view missing data completion framework using sMRI and FDG-PET ROI-based features. Their method achieved a classification accuracy of $0.836$ for classifying CN versus AD, even with $50\%$ missing PET.
Similarly, Ye and colleagues \cite{ye2022pairwise} proposed a GAN with an attention layer to generate missing FDG-PET from available MRI features, resulting in a classification accuracy of $0.914\pm0.19$. 
In the same way, Zhang et al. \cite{zhang2024pyramid} employed a GAN model based on a pyramidal attention mechanism for imputing PET data from sMRI. Then, the generated PET images were fused with the sMRI at pixel-level, and then the same generator used for PET synthesis was employed for feature extraction, extracting a latent representation that fed a dense layer, reaching an accuracy of $0.934$.
Zhang et al. \cite{zhang2022bpgan} developed a GAN-based model, named BPGAN, for generating FDG-PET data from sMRI. They created a 3D multiple convolution U-Net (MCU) generator for PET images, and then they fused the sMRI and the synthetic PET volumes concatenating the data at the pixel-level. Finally, they employed a ResNet-18 for classification achieving an accuracy of $0.981$ in discriminating AD from CN.
Gao et al. \cite{gao2023multimodal} proposed a multi-level guided GAN (MLG-GAN) and a multimodal transformer (Mul-T) for synthesizing PET and T2-MRI images from sMRI and extracting features for classification from the three modalities, respectively. The MLG-GAN module generated data using voxel-, feature-, and task-level information for a more accurate generation. The Mul-T achieved competitive performance by integrating local and global information from all modalities reaching an accuracy of $0.944$.
Tu and colleagues \cite{tu2024multimodal} developed a consistent manifold projection GAN (CMPGAN) model for imputing missing PET from sMRI. The CMPGAN integrated a manifold projection operator to map the data in a low dimensional space and a distribution distance measure to ensure that no gradient disappeared during the generator training, hence creating more reliable data. Then, a multilevel multimodal fusion diagnosis network (MMFDN) was developed to diagnose AD by fusing together the synthetic PET and sMRI images. They obtained a final performance in CN versus AD of $0.980\pm0.003$.
Finally, Gao and colleagues \cite{gao2021task} presented a task-induced pyramid and attention generative adversarial network (TPA-GAN) to generate missing FDG-PET data from sMRI, achieving an accuracy of $0.927$. 
%However, the model was trained and tested on two independent databases with different acquisition protocols, which on one side allowed to evaluate the generalizability of the model, but on the other could bias the results due to the different data source.
Moreover, it still lacked the ability to reconstruct sMRI from FDG-PET data. 
While these approaches demonstrated excellent results, their framework was limited in handling arbitrary missing modalities.
%, allowing only the completion of missing PET from available sMRI and FDG-PET features.

As mentioned before, the trained model performance was subsequently assessed on Task 2.
Only the $7\%$ of the dataset shared all the modalities, making it a challenging scenario. Nevertheless, our framework achieved an average accuracy of $0.711\pm0.01$ for the independent test sets.
%, assessing better performance with respect to the baseline methods, $0.690\pm0.01$ and $0.698\pm0.01$ for SVM and random forest with mean imputation strategy, and $0.688\pm0.01$ and $0.694\pm0.0$ with zero imputation approach, respectively. 
Although it did not outperform methods specifically trained on this particular task, it demonstrated competitive results. Other approaches have focused on different input data and missing data management to solve this task. For instance, Ritter and colleagues \cite{ritter2015multimodal} proposed simple machine learning methods with tabular features and limited imputation techniques, achieving an accuracy of $0.670$ to stratify MCI subjects.
Cai et al. \cite{cai2018deep} used a GAN to impute missing PET images from sMRI scans, achieving an accuracy of $0.657$ for discriminating MCInc vs MCIc patients. Furthermore, Zhou and colleagues \cite{zhou2019latent} proposed a framework for projecting original features into a latent representation, resulting in an accuracy of $0.743$ in case of missing the $51$\% of PET data. Gao et al. \cite{gao2021task} applied the approach already discussed also to address the classification Task 2, reaching an accuracy score of $0.753$. Gao et al. \cite{gao2023multimodal} employed the same framework described in Task 1 of the MCI conversion task. They initially pretrained the models on AD detection, and then they refined the weights on the MCI task, obtaining a final accuracy of $0.778$. Similarly, Tu and colleagues \cite{tu2024multimodal} trained and tested their model described for Task 1 in Task 2. They achieved an accuracy of $0.923\pm0.01$ in the MCI conversion task.

Despite the promising results obtained by the state-of-the-art generative models for missing data imputation, it is essential to acknowledge their shared limitation: they all relied on having sMRI as a prerequisite to impute PET data, and very few included genetics, T2-MRI, or rs-fMRI information in their analyses, which have instead been demonstrated as relevant biomarkers for AD \cite{pini2021breakdown, franzmeier2020functional}. 
In contrast, our framework does not necessitate a shared modality across all subjects. The exploitation of two cGANs during the generation phase enables to produce the missing latent rs-fMRI and/or SNPs from sMRI. In addition, this process is applicable bidirectionally, allowing the generation of sMRI latent features from either SNPs or rs-fMRI as well. The versatility of this approach opens the way to its generalization to additional modalities through training distinct cGANs and integrating the resulting generators into the complete classification framework. 
Furthermore, our model's performance confirms that generative models allow to obtain realistic data and learn nonlinear mappings across the different acquisitions, achieving competitive prediction accuracy also with a substantial proportion of missing modalities.
Supplementary Table 7 shows the performance of the proposed model compared with the state-of-the-art competitors allowing multimodality and missing data management for both Task 1 and Task 2.

\subsection{Interpretability analyses}
\label{SS2}
Interpretability analysis was performed relying on IG, followed by group-based statistical analysis for the sake of validation. More in detail, for each input modality and each task, relevance maps were extracted and analyzed from a qualitative and quantitative point of view.
Interestingly, among the state-of-the-art methods for multimodal AD or MCI conversion detection, only a few works introduced interpretability analysis. Venugopalan et al. \cite{venugopalan2021multimodal} proposed an occlusion-based approach to extract the feature relevance. Ye and colleagues \cite{ye2022pairwise} relied on an attention layer in their generation module for obtaining the feature importance. Zhou et al. \cite{zhou2019latent} exploited the weights learned for their latent representation learning to derive the input features' importance. Zhang and colleagues \cite{zhang2024pyramid} employed 3D Grad-CAM for extracting the most relevant regions from the fused MRI-PET images that were of interest for their model. While Gao et al. \cite{gao2023multimodal} extracted the spatial attention maps extracted from the attention mechanism. In general, these works found brain regions, such as the hippocampus and amygdala, as the most important among the others, which was in line with our findings.
In our results, the interpretability analysis revealed similarities between the MCIc and the AD phenotypes, as well as between the MCInc and CN, while highlighting significant differences between the CN and AD, as expected, but also between MCIc and MCInc which are of high interest since they would allow to identify early AD biomarkers. 

Regarding the sMRI data, our findings highlighted that sub-cortical regions carried more informative and relevant information for the final classification compared to cortical areas. The hippocampus resulted as the most critical region for all classes under analysis, alongside the amygdala. In AD and MCIc patients, such regions exhibited negative attribution values, suggesting a decrease in GM probability, as opposed to positive values in CN and MCInc subjects. Of interest, the most significant differences in the relevance assigned to such regions were found between MCIc and MCInc. These findings were in line with the well-known literature and established hallmarks of AD. Indeed, the pathological process initially affects the hippocampus and amygdala before extending to other nearby structures \cite{lehericy1994amygdalohippocampal, barnes2009meta}. Focusing on MCI, Liu and collegaues\cite{liu2010analysis} demonstrated that MCIc patients exhibited higher atrophy in the hippocampus and amygdala compared to MCInc and CN subjects, which aligns well with our findings. Regarding the caudate region, Sodums et al. \cite{sodums2020negative} and Persson and colleagues \cite{persson2018finding} highlighted a negative correlation between caudate and hippocampus volumes in healthy subjects, and, in addition, the second work also showed that patients with AD and non-specified dementia have a larger caudate volume compared to non-dementia subjects. However, this is still debated in literature with other works suggesting that the caudate is susceptible to atrophy, resulting in a reduction of the GM volume of this region in AD patients \cite{rombouts2000unbiased,jiji2013segmentation,frisoni2002detection}.
The cortical areas had a generally lower relevance to the classification for all the phenotypes. The parahippocampal gyrus (anterior division) and the occipital fusiform gyrus showed the highest attribution scores, with positive values assigned to CN. This was in line with Wang and colleagues \cite{wang2015voxel} and Liu et al. \cite{liu2010analysis} findings which revealed smaller volumes in the anterior parahippocampal area in AD, MCIc, and MCInc subjects compared to CN individuals.
Additionally, Dai et al. \cite{dai2012discriminative} highlighted the importance of the occipital fusiform gyrus for AD classification. As for the other relevant brain regions, the precuneus and fusiform cortices showed a reduction of the GM volume in our results, coherent with what is known from the literature \cite{apostolova2007three,karas2007precuneus,moller2013different}.
Overall, our findings on sMRI revealed that the relevance scores were sensible to an increase in neurodegeneration in some focal brain areas in AD and MCIc compared with CN and MCInc, which aligns well with the literature findings.

Based on rs-fMRI attributions, five functional RSNs emerged as highly relevant for the final classification of both tasks, namely SM, DM, CB, VI, and CC.
The DM has a central role in information integration and processing, and its involvement in AD is well-known in the current literature, with several studies consistently demonstrating that this is the first RSN to be affected by abnormal protein aggregation \cite{zhan2016network, jones2016cascading}. In our results, we did not retrieve intra-network connections in the DM mode. However, for the MCIc subjects, a negative relevance score, suggesting a decreased FC, was found for inter-network connections between DM and visual/sensorimotor areas (DM-VI, DM-SM), while a positive inter-network connection was highlighted between DM and CC. Conversely, AD subjects revealed a positive relevance for a few connections between the DM and VI and between the DM and the CC, while one negative connection was found between DM and SM. These patterns are not commonly reported in the literature but deserve further investigation since they could represent compensation connectivity patterns. 
Among the other RSNs, the SM was the most present and relevant, showing negative relevance scores for AD and MCIc for both the intra-connections, involving post/para central and partial gyri, and inter-connections with the other relevant RSNs. On the contrary, for the CN and, importantly, for the MCInc, the SM showed positive relevance, leading to significant differences for both intra- and inter-network connections with the most severe groups (AD and MCIc). The lowest \textit{p}-values were found when comparing MCInc vs AD as well as CN vs AD over the different SM intra-network connections. Inter- and intra-network SM connections were indeed demonstrated to be affected by the AD pathology, generally showing an overall decreased connectivity in the later stages of the disease \cite{wang2015aberrant,damoiseaux2012functional,zhan2016network}. Moreover, Albers et al. \cite{albers2015interface} showed that many pyramidal and extrapyramidal motor impairments affect a substantial portion of AD patients, even at an early stage of the disease, and progressively worsen along with cognitive impairment, reflecting a possible decreased connectivity in the SM network.
Interestingly, the VI RSN was involved in the classification of the different AD stages. In detail, positive relevance was assigned to AD subjects in both intra- and inter-connections (VI-DM and VI-CC), while a different pattern was found for MCInc, which showed negative relevance in almost the same ICs. Of note, the MCIc did not show high involvement of the VI intra-connections, while few negative inter-connections were found between VI and the other relevant RSN (SM, CC, and DM). This revealed another difference between AD and its prodromal stages, indicating a decrease in VI connectivity in the MCI stage, particularly evident for the MCInc, which then converts into hyperconnectivity in the most severe stage, full-blown dementia. The damage in VI due to AD was previously discussed by Zhan and colleagues \cite{zhan2016network}. In detail, Albers et al. \cite{albers2015interface} found that subgroups of AD patients have concomitant eye diseases, and some visual functions are impaired, which could be caused by impairments in the VI RSN. Moreover, recent studies demonstrated a hyperconnectivity pattern in the most severe stages of the disease present in the VI network \cite{penalba2023increased}. Hence, further investigation would be needed to elucidate the involvement of VI RSN in the AD continuum.
%Finally, the inter-network connections between the CB and the SM resulted negatively relevant, hence affected by the disease with a decreased connectivity in both AD and MCIc, while the same was not recorded for the MCInc, unraveling a possible impact on this RSN in a later stage compared to the others.

Regarding the genetics impact, the significant biological processes for MCInc, MCIc, and AD were derived from the genes annotated starting from the most significant SNPs resulting from the interpretability analysis on the two tasks. 
Biological processes related to the regulation of endocytosis were found both in AD and MCIc. Previous studies highlighted how endocytosis is strongly related to AD \cite{zadka2024endocytosis}. Endocytosis is a pathway that, with all the components, proteins, and membrane organization and modulations related to it, plays an important role in AD pathology since it is involved in the trafficking and clearance of A$\beta$ proteins \cite{zadka2024endocytosis}.
Moreover, particularly in MCIc, we found biological processes related to cholesterol, like \textit{cholesterol transport} and \textit{sterol transport}, which, on a deeper analysis, were shown as possibly being associated with the AD continuum \cite{zadka2024endocytosis}. Excess cholesterol deposit in the brain was demonstrated to be related to an increase of A$\beta$ plaques and amyloid cascade leading to synaptic plasticity annihilation and promotion of tau phosphorylation, hence contributing to the risk of AD pathogenesis, possibly in an early phase \cite{shobab2005cholesterol,michikawa2003role}. Additionally, for the MCIc individuals, a biological process related to A$\beta$ was highlighted, the \textit{negative regulation of amyloid precursor protein catabolic process} that is related to AD \cite{dovrolis2022unlocking,mills1999regulation}. In relation to A$\beta$ proteins, some biological pathways related to the regulation of amide were found in MCIc individuals. A previous study found that fatty acid amides were strongly associated with A$\beta$ and the hippocampal volume \cite{kim2019primary}. Regarding the particular genes, APOE, CLU, PICALM, APOA2, ABCA7, and BIN1 were among the most frequent in the biological processes. They all have a relevant impact in the development of AD \cite{scheltens2021alzheimer,hu2017analyzing,ma2015serum}. 
In particular, APOE is notably the major risk factor of AD \cite{kim2009role,scheltens2021alzheimer} and is mainly expressed in both the brain and the liver. More in-depth, the ApolipoproteinE is a ligand receptor-mediated endocytosis of lipoprotein particles \cite{kim2009role} and is the major cholesterol transport and other lipids in the brain \cite{holtzman2012apolipoprotein,wang2021regulation}. This gene is hence strongly related to one of the most important biomarkers for AD, namely A$\beta$ plaques containing A$\beta$ peptides and the neurofibrillary tangles containing hyperphosphorylated tau proteins \cite{munoz2019understanding}.
TREM2 is another gene-coding protein with a key role in AD progression, and the rare variant of TREM2 R47H has a high-risk factor for AD comparable to the strongest biomarker APOE. TREM2 is a transmembrane receptor on the microglia that, in the condition of A$\beta$ aggregation, rapidly migrates closer to the aggregation, transforming them and promoting the phagocytosis and clearance of A$\beta$ formation. Hence, one of the roles of TREM2 is the regulation of microglia recruitment closer to A$\beta$ formations for the uptake and degradation of them \cite{zhong2019soluble}.
ABCA7 is one of the most important risk genes for AD \cite{de2019role} that mainly regulates the processes related to cholesterol and the processing of A$\beta$ proteins \cite{dib2021role}. ABCA7-expressed proteins have a relevant role in the control of cholesterol metabolism, and then the cholesterol has a strong influence in the regulation of A$\beta$ synthesis \cite{dib2021role,martins2009cholesterol}. Additionally, some studies reported also that different variants of ABCA7 are associated with an increase of A$\beta$ deposition in MCI patients rather than AD \cite{apostolova2018associations}.
Similarly, APOA2 is strictly related to cholesterol and HDL since it transports the cholesterol to the liver and is an important component for the formation of HDL \cite{ma2015serum}.
As for the other relevant genes, PICALM is involved in endocytic-related processes, suited for the production, modulation, and clearance of A$\beta$ complexes \cite{zadka2024endocytosis,xu2015role}. Also, for this gene, interactions with APOE have been demonstrated \cite{jun2010meta}. 
Along with PICALM, the CLU variant was identified as an important biomarker for AD \cite{harold2009genome} since it is implicated in white matter integrity, membrane recycling, and lipid transportation \cite{braskie2011common}.
Finally, Yu et al. \cite{yu2015association} and Chapuis and colleagues \cite{chapuis2013increased} found that BIN1 is a gene strictly related to A$\beta$ and $\tau$ pathologies, and presumably regulates the APOE \cite{scheltens2021alzheimer}, hence assuming high relevance in AD.

\subsection{Main contributions and outcomes}
We proposed an interpretable, flexible, and generative framework for multimodal AD detection and MCI conversion prediction. The main contributions of this work are as follows: (i) reaching the state-of-the-art of multimodal and generative models in the classification of CN vs AD; (ii) reaching competitive performance in the segregation of MCInc/MCIc phenotypes using a pretrained framework; (iii) managing missing data introducing a generation module in the latent space allowing to impute missing modalities relaxing the constraint of having at least one modality shared by the whole cohort, and (iv) proposing an interpretability analysis based on IG for extracting biological information.

The complementary analysis of three different input modalities allowed to uncover the disease signatures at multiple levels of analysis, providing complementary yet interdependent information. Besides some results being in agreement with what is known from the literature, providing evidence of the trustability of the outcomes of the proposed analysis, other findings suggest the involvement of additional mechanisms that could contribute to elucidate the mechanisms underlying the pathogenesis and progression of the disease and would deserve further investigation.

The main findings concern the presence of atrophy, particularly involving the hippocampus in later stages (MCIc and AD), that is well known in the literature, FNC modulation, particularly in the information processing-related RSN (SM and VI) as well as SNPs mutations in phenotype-specific genes. Regarding FNC, for SM negative relevance scores were recorded, hence suggesting a decreased inter- and intra-connectivity, in both the MCIc and the AD stages, while for VI a different pattern was observed for MCInc/MCIc/AD subjects. In particular, negative attributions were recorded in the MCInc stage, with reduced relevance in the MCIc, while positive attribution values characterized the AD, suggesting a compensation mechanism leading to a hyperconnectivity in such RSN in the later stages of the disease. Such brain modulations were present along with the mutation of relevant SNPs linked to the involvement of different biological processes related to endocytosis, cell and membrane transportation and organization, amyloid and cholesterol regulation for the AD and MCIc subjects, while mostly related to the T cell processes, white blood cells, and immune response pathways for the MCInc.

\subsubsection{Limitations and future works}
In what follows, the main open issues will be briefly summarized, paving the way for further research. First, concerning the MCI task, in this work, our aim was to test the model generalizability by straightforwardly testing it on the MCI cohort. Training and fine-tuning on the specific task would allow obtaining better classification results and will be the object of future analysis. Being the model intrinsically multimodal and easily flexible, additional channels could be included, such as the clinical information and PET imaging, which have been demonstrated to be highly discriminative for the disease \cite{knopman2021alzheimer}, as well as advanced structural connectivity metrics derived from diffusion MRI which could help to shed light on early disease signatures. On top of this, it would be interesting to explore different phenotype stratification tasks, either proposing a multiclass classification or defining biologically homogeneous groups following the A/T/N (amyloid, tau, neurodegeneration) system which has been attracting increasing attention in recent years \cite{jack2016t}. Finally, concerning the interpretability analysis, we strongly point toward deeper exploitation of XAI methods to open the way to not yet studied associations or mechanisms that are captured by the model. The next step would be the post-hoc validation of the XAI outcomes pursuing the robustness, reliability, and trustfulness of the interpretability analysis outcomes. Last but not least, the analysis of the associations between the different input features would support the interpretation of the results as well as further analysis for elucidating causal relations across the investigated underlying mechanisms.

\section{Conclusions}
In this work, we presented a multimodal, flexible, generative, and interpretable DL-based framework for AD detection and MCI conversion classification. Neuroimaging (structural and functional features) and genetics data were used to address these classification tasks.
The generation of missing modalities in the latent space using four pre-trained generators of two different cGANs allowed to obtain competitive classification performance in both tasks.
The application of an interpretability method yielded our model to be interpretable extracting the relevance of each input feature and revealing the most important ones for each class, highlighting disease structural, functional, and genetic signatures and opening the way to further analyses.

\section*{Declaration of competing interest}
The authors declare that they have no known competing financial interests or personal relationships that could have appeared to influence the work reported in this paper.

\section*{Acknowledgments}
Data collection and sharing for this project was funded by ADNI (NIH Grant U01 AG024904) and DOD ADNI (DOD award number W81XWH-12-2-0012). ADNI is funded by the NIA, the NIBIB, and through generous contributions from the following: AbbVie, Alzheimer’s Association; Alzheimer’s Drug Discovery Foundation; Araclon Biotech; BioClinica, Inc.; Biogen; Bristol-Myers Squibb Company; CereSpir, Inc.; Cogstate; Eisai Inc.; Elan Pharmaceuticals, Inc.; Eli Lilly and Company; EuroImmun; F. Hoffmann-La Roche Ltd and its affiliated company Genentech, Inc.; Fujirebio; GE Healthcare; IXICO Ltd.; Janssen Alzheimer Immunotherapy Research \& Development, LLC.; Johnson \& Johnson Pharmaceutical Research \& Development LLC.; Lumosity; Lundbeck; Merck \& Co., Inc.; Meso Scale Diagnostics, LLC.; NeuroRx Research; Neurotrack Technologies; Novartis Pharmaceuticals Corporation; Pfizer Inc.; Piramal Imaging; Servier; Takeda Pharmaceutical Company; and Transition Therapeutics. The CIHR is providing funds to support ADNI clinical sites in Canada. Private sector contributions are facilitated by the FNIH (\url{www.fnih.org}). The grantee organization is the NCIRE, and the study is coordinated by the Alzheimer’s Therapeutic Research Institute at USC. ADNI data are disseminated by the Laboratory for Neuro Imaging at USC. 

The study was also funded by NIH grant RF1AG063153 and NSF grant \#2112455, as well as, Fondazione CariVerona (Bando Ricerca Scientifica di Eccellenza 2018, EDIPO project, num. 2018.0855.2019) and MIUR D.M. 737/2021 ``AI4Health: empowering neurosciences with eXplainable AI methods''.

\bibliographystyle{unsrt}
\bibliography{arxiv_JNE_GD.bib}

\end{document}

\maketitle

\section*{Materials and Methods}
\subsection*{Preprocessing quality control of MRI}
A thorough quality control (QC) was performed to retain scans with good normalization to the standard MNI space, which involved discarding sMRI and fMRI images that exhibited low correlation with individual and/or group-level masks. In this spatial correlation process, we first calculated subject-level masks using the subject MRI scans (only the first volume in case of fMRI) by setting the brain voxels to 1 if the values of these voxels were greater than $80\%$ of the average value across whole-brain voxels, and 0 otherwise. Next, after computing the subject-level masks, we calculated a group mask by setting the voxels to 1 for which at least $70\%$ of the subject-level masks had a value of 1. Lastly, we examined the spatial correlations of the subject and group level masks and retained subjects that showed a correlation value greater than $0.85$. Additionally, for fMRI, scans with larger head motion parameters ($>3^\circ$ rotations and $>3$ mm translations) were discarded.

\subsection*{Training scheme additional information}
The weights of the CNNs and classifier, used in both bi-modal models and the final framework, were initialed as follows: the 3D CNN for sMRI, the 1D CNN for SNPs, and the classifier using the Xavier Uniform distribution, while Xavier Normal distribution was used for the 1D CNN for rs-fMRI. The weights of the generators and discriminators were initialized following a Xavier Normal distribution for both the sMRI-fMRI-cGAN and sMRI-SNPs-cGAN. The weights initialization was independent between Step 1 and Step 2.

No data leakage was present between the training Step 1 (bi-modal models and cGANs) and the training Step 2 (full multimodal framework) since Step 2 involved end-to-end training using subjects with no missing modalities (i.e., using input data for these subjects only) and generative models trained in Step 1b for deriving latent features for subjects with missing modalities only. The full cohort of subjects was not used to train the bi-modal models and corresponding cGANs (Step 1a and 1b), only relying on those subjects sharing the modalities under analysis (sMRI and rs-fMRI for sMRI-fMRI-NN/cGAN, and sMRI and SNPs for sMRI-SNPs-NN/cGAN).
    
\section*{Results}
\subsection*{Classification comparison with baseline approaches}
Table \ref{tab:baseline_comparison} summarizes the performance comparison of the proposed framework with the baseline methods.
Regarding Task 1, AD detection, our proposed framework showed higher performance with respect to the baseline methods used for comparison. Additionally, the ensemble SVM model with mean imputation (SVM $\mu$) and SVM with zero imputation (SVM 0) showed higher performance in Task 1 than random forest (considering both imputation methods). On the other hand, random forest with mean imputation (RF $\mu$) and zero imputation (RF 0) strategies achieved higher performance values in Task 2 with respect to SVM approaches. Additionally, in Task 2, the proposed model outperformed in terms of accuracy the baseline methods, but both random forest approaches achieved a precision value higher than the proposed framework.

\begin{table}[!ht]
    \centering
    \caption{Performance comparison of the proposed framework with two baseline methods, SVM and random forest (RF), and two imputation strategies, mean imputation ($\mu$) and zero imputation ($0$).}
    \begin{adjustbox}{width=\linewidth}
    \begin{tabular}{l c c c c c c c}
    \toprule
        ~ & \multicolumn{3}{c}{Task 1} & \multicolumn{3}{c}{Task 2} \\[3pt]
        ~ & Accuracy & Precision & Recall & & Accuracy & Precision & Recall \\[3pt]
        \cline{2-4} \cline{6-8} \\[-1.5ex]
        SVM $\mu$ & $0.699\pm0.01$ & $0.852\pm0.09$ & $0.149\pm0.05$ & & $0.690\pm0.01$ & $0.399\pm0.16$ & $0.074\pm0.06$ \\[3pt]
        SVM $0$ & $0.689\pm0.02$ & $0.900\pm0.08$ & $0.105\pm0.06$ & & $0.688\pm0.01$ & $0.325\pm0.15$ & $0.041\pm0.06$ \\[3pt]
        RF $\mu$ & $0.663\pm0.01$ & $0.533\pm0.45$ & $0.018\pm0.02$ & & $0.698\pm0.01$ & $0.804\pm0.17$ & $0.028\pm0.01$ \\[3pt]
        RF $0$ & $0.663\pm0.01$ & $0.800\pm0.40$ & $0.015\pm0.01$ & & $0.694\pm0.01$ & $0.836\pm0.14$ & $0.015\pm0.01$ \\[3pt]
        Our & $0.926\pm0.02$ & $0.910\pm0.05$ & $0.876\pm0.03$ & & $0.711\pm0.01$ & $0.558\pm0.03$ & $0.610\pm0.03$ \\[3pt]
        \bottomrule
    \end{tabular}
    \end{adjustbox}
    \label{tab:baseline_comparison}
\end{table}

\begin{table}[!ht]
    \centering
    \caption{In this Table are reported the correspondences acronym - full name of the sMRI brain regions under analysis.}
    \begin{adjustbox}{width=\textwidth}
    \begin{tabular}{ l  l  l  l  l  l }
        \toprule
        \makecell[c]{Acronym} & \makecell[c]{Full name} & \makecell[c]{Acronym} & \makecell[c]{Full name} & \makecell[c]{Acronym} & \makecell[c]{Full name} \\
        \midrule
        Ins & Insular Cortex & OFG & Occipital Fusiform Gyrus & TP & Temporal Pole \\
        COpC & Central Opercular Cortex & ScC & Subcallosal Cortex & POpC & Parietal Operculum Cortex \\
        CGp & Cingulate Gyrus, posterior division & Thl & Thalamus & PcC & Precuneous Cortex \\
        Cau & Caudate & FOC & Frontal Orbital Cortex & Put & Putamen \\
        PhGa & Parahippocampal Gyrus, anterior division & Pall & Pallidum & PaGp & Parahippocampal Gyrus, posterior division \\
        Hipp & Hippocampus & LG & Lingual Gyrus & Amy & Amygdala \\
        TFCp & Temporal Fusiform Cortex, posterior division & Acc & Accumbens & TOF & Temporal Occipital Fusiform Cortex \\
        \bottomrule
    \end{tabular}
    \end{adjustbox}
    \label{tab:acronyms_sMRI_table}
\end{table}

\begin{table*}[!ht]
    \centering
    \caption{In this Table are reported the $53$ ICs present in the sFNC matrices with corresponding brain region name, network at which it belongs to, and spatial location in the brain along X, Y, and Z axis. SC=Sub-cortical; AU=Auditory; SM=sensorimotor; VI=visual; CC=cognitive-control; DM=default-mode; and CB=cerebellar. In \textit{italic} are highlighted the brain regions present in the fMRI connectograms of Results section.}
    \begin{adjustbox}{width=\textwidth}
    \begin{tabular}{ c  l  c  c  c  c  c  c  l  c  c  c  c }
        \toprule
        \makecell[c]{IC ID} & \makecell[c]{Brain region} & \makecell[c]{Network} & \makecell[c]{X} & \makecell[c]{Y} & \makecell[c]{Z} & & \makecell[c]{IC ID} & \makecell[c]{Brain region} & \makecell[c]{Network} & \makecell[c]{X} & \makecell[c]{Y} & \makecell[c]{Z}\\
        \midrule
        1 & Caudate & SC & 6.5 & 10.5 & 5.5 & & 26 & Inferior parietal lobule & CC & 45.5 & -61.5 & 43.5 \\
        2 & Subthalamus/hypothalamys & SC & -2.5 & -13.5 & -1.5 & & 27 & Insula & CC & -30.5 & 22.5 & -3.5 \\
        3 & Putamen & SC & -26.5 & 1.5 & -0.5 & & 28 & \textit{Superior medial frontal gyrus} & CC & -0.5 & 50.5 & 29.5 \\
        4 & Caudate & SC & 21.5 & 10.5 & -3.5 & & 29 & Inferior frontal gyrus & CC & -48.5 & 34.5 & -0.5 \\
        5 & Thalamus & SC & -12.5 & -18.5 & 11.5 & & 30 & Right inferior frontal gyrus & CC & 53.5 & 22.5 & 13.5 \\
        6 & Superior temporal gyrus & AU & 62.5 & -22.5 & 7.5 & & 31 & Middle frontal gyrus & CC & -41.5 & 19.5 & 26.5 \\
        7 & Middle temporal gyrus & AU & -42.5 & -6.5 & 10.5 & & 32 & Inferior parietal lobule & CC & -53.5 & -49.5 & 43.5 \\
        8 & \textit{Postcentral gyrus} & SM & 56.5 & -4.5 & 28.5 & & 33 & \textit{Left inferior parietal lobule} & CC & 44.5 & -34.5 & 46.5 \\
        9 & \textit{Left postcentral gyrus} & SM & -38.5 & -22.5 & 56.5 & & 34 & Supplementary motor area & CC & -6.5 & 13.5 & 64.5 \\
        10 & \textit{Paracentral lobule} & SM & 0.5 & -22.5 & 65.5 & & 35 & Superior frontal gyrus & CC & -24.5 & 26.5 & 49.5 \\
        11 & \textit{Right postcentral gyrus} & SM & 38.5 & -19.5 & 55.5 & & 36 & Middle frontal gyrus & CC & 30.5 & 41.5 & 28.5 \\
        12 & \textit{Superior parietal lobule} & SM & -18.5 & -43.5 & 65.5 & & 37 & Hippocampus & CC & 23.5 & -9.5 & -16.5 \\
        13 & \textit{Paracentral lobule} & SM & -18.5 & -9.5 & 56.5 & & 38 & \textit{Left inferior parietal lobule} & CC & 45.5 & -61.5 & 43.5 \\
        14 & \textit{Precentral gyrus} & SM & -42.5 & -7.5 & 46.5 & & 39 & Middle cingulate cortex & CC & -15.5 & 20.5 & 37.5 \\
        15 & \textit{Superior parietal lobule} & SM & 20.5 & -63.5 & 58.5 & & 40 & Inferior frontal gyrus & CC & 39.5 & 44.5 & -0.5 \\
        16 & \textit{Postcentral gyrus} & SM & -47.5 & -27.5 & 43.5 & & 41 & Middle frontal gyrus & CC & -26.5 & 47.5 & 5.5 \\
        17 & \textit{Calcarine gyrus} & VI & -12.5 & -66.5 & 8.5 & & 42 & \textit{Hippocampus} & CC & -24.5 & -36.5 & 1.5 \\
        18 & \textit{Middle occipital gyrus} & VI & -23.5 & -93.5 & -0.5 & & 43 & Precuneus & DM & -8.5 & -66.5 & 35.5 \\
        19 & Middle temporal gyrus & VI & 48.5 & -60.5 & 10.5 & & 44 & \textit{Precuneus} & DM & -12.5 & -54.5 & 14.5 \\
        20 & \textit{Cuneus} & VI & 15.5 & -91.5 & 22.5 & & 45 & \textit{Anterior cingulate cortex} & DM & -2.5 & 35.5 & 2.5 \\
        21 & \textit{Right middle occipital gyrus} & VI & 38.5 & -73.5 & 6.5 & & 46 & Posterior cingulate cortex & DM & -5.5 & -28.5 & 26.5 \\
        22 & \textit{Fusiform gyrus} & VI & 29.5 & -42.5 & -12.5 & & 47 & \textit{Anterior cingulate cortex} & DM & -9.5 & 46.5 & -10.5 \\
        23 & \textit{Inferior occipital gyrus} & VI & -36.5 & -76.5 & -4,5 & & 48 & \textit{Precuneus} & DM & -0.5 & -48.5 & 49.5 \\
        24 & \textit{Lingual gyrus} & VI & -8.5 & -81.5 & -4.5 & & 49 & Posterior cingulate cortex & DM & -2.5 & 54.5 & 31.5 \\
        25 & \textit{Middle temporal gyrus} & VI & -44.5 & -57.5 & -7.5 & & 50 & \textit{Cerebellum} & CB & -30.5 & -54.5 & -42.5 \\
         &  &  &  &  & & & 51 & \textit{Cerebellum} & CB & -32.5 & -79.5 & -37.5 \\
         &  &  &  &  & & & 52 & Cerebellum & CB & 20.5 & -48.5 & -40.5 \\
         &  &  &  &  & & & 53 & \textit{Cerebellum} & CB & 30.5 & -63.5 & -40.5 \\
        \bottomrule 
    \end{tabular}
    \end{adjustbox}
    \label{tab:ICs_table}
\end{table*}

\subsection*{Group-based statistical IG analysis}
\subsubsection*{Neuroimaging modalities}
\begin{figure*}[!ht]
    \centering
    \includegraphics[width=\textwidth]{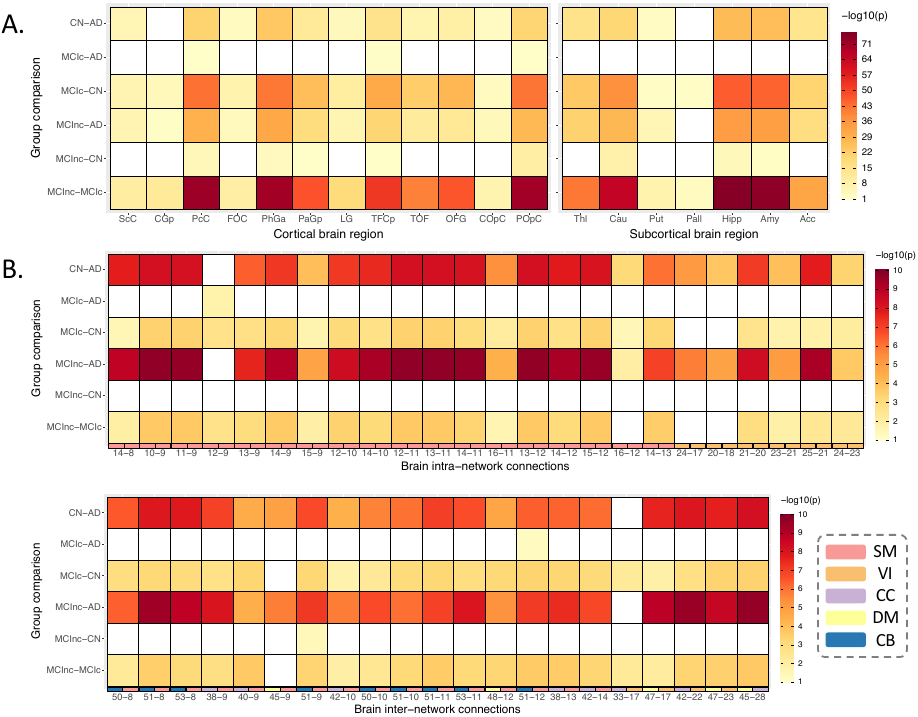}
    \caption{Overview of the neuroimaging statistical analysis. A. sMRI-IG Wilcoxon test results for each group comparison and each brain region, divided into cortical and subcortical regions; B. fMRI-IG Wilcoxon test results for each group comparison for each relevant brain connection divided into extra- (upper) and intra-network (lower) connectivity. The \textit{p}-values for B. and C. are reported in negative logarithmic scale, dark red the most significant. White cells represent no statistical significance.}
    \label{fig:post_hoc_neuroimaging}
\end{figure*}

Supplementary Figure \ref{fig:post_hoc_neuroimaging}A shows the statistical analysis results of sMRI data. The heatmap reports the Bonferroni corrected \textit{p}-values ($6$ pairwise comparisons) in logarithmic scale (yellow/red correspond to higher/lower \textit{p}-value, and white means non-significant). While supplementary Figure \ref{fig:post_hoc_neuroimaging}B shows the statistical analysis results of rs-fMRI data.

\begin{table*}[!ht]
    \centering
    \caption{Most significant biological processes for AD patients obtained from the analysis of the most relevant SNPs with positive IG attributions. In this Table are shown the biological processes, their raw \textit{p}-value, and the overlap genes. $^*$ indicates statistically significant biological processes after Bonferroni correction.}
    \begin{adjustbox}{width=\textwidth}
    \begin{tabular}{ l  l  l  l }
        \toprule
        \makecell[c]{Biological processes} & \makecell[c]{GO term} & \makecell[c]{p-value} & \makecell[c]{Overlap genes}\\
        \midrule
        Intracellular transport$^*$ & GO:0046907 & \makecell[l]{$0.000044$\\(p$_{bonf}=0.0225$)} & \makecell[l]{BLOC1S3; KLC3; TOMM40; PICALM; MME; NUP88; TMEM106B;\\ NSF; NUP43; BIN1; RAB12; APOE; SORL1; CLU} \\[10pt]
        Regulation of protein-containing complex assembly$^*$ & GO:0043254 & \makecell[l]{$0.000080$\\(p$_{bonf}=0.0413$)} & \makecell[l]{PTK2B; MARK4; FNIP2; PLCG2; BIN1; APOE; SORL1; TREM2;\\ CLU} \\[10pt]
        Establishment of localization in cell$^*$ & GO:0051649 & \makecell[l]{$0.000086$\\(p$_{bonf}=0.0446$)} & \makecell[l]{BLOC1S3; KLC3; TOMM40; PICALM; MME; NUP88; TMEM106B;\\ NSF; NUP43; BIN1; RAB12; APOE; SORL1; NECTIN2; CLU} \\[10pt]
        Cell development$^*$ & GO:0048468 & \makecell[l]{$0.000091$\\(p$_{bonf}=0.0470$)} & \makecell[l]{BLOC1S3; KLC3; PICALM; PTK2B; ALDH1A2; TNXB; TMEM106B;\\ FCER1G; OOSP2; PLCG2; IGSF23; BIN1; ERCC2; RELB; APOE;\\ ETV1; NECTIN2; TREM2; CLU} \\[15pt]
        Cell projection organization & GO:0030030 & $0.000253$ & \makecell[l]{BLOC1S3; KLC3; PICALM; PTK2B; TNXB; TMEM106B; MARK4;\\ CDH13; APOE; ETV1; MTSS2; NECTIN2; TREM2} \\[10pt]
        Plasma membrane bounded cell projection organization & GO:0120036 & $0.000253$ & \makecell[l]{BLOC1S3; KLC3; PICALM; PTK2B; TNXB; TMEM106B; MARK4;\\ CDH13; APOE; ETV1; MTSS2; NECTIN2; TREM2} \\[10pt]
        Membrane organization & GO:0061024 & $0.000253$ & \makecell[l]{TOMM40; PICALM; RABEP1; NSF; APOA2; BIN1; CR1; RAB12;\\ APOE; MTSS2; NECTIN2; TREM2; CLU} \\[10pt]
        Developmental maturation & GO:0021700 & $0.000417$ & \makecell[l]{BLOC1S3; PICALM; PTK2B; ALDH1A2; SLC24A4; OOSP2; ERCC2} \\[5pt]
        Positive regulation of endocytosis & GO:0045807 & $0.000417$ & \makecell[l]{FCER1G; PLCG2; APOA2; BIN1; APOE; TREM2; CLU} \\[5pt]
        Regulation of vesicle-mediated transport & GO:0060627 & $0.000427$ & \makecell[l]{PICALM; NSF; FCER1G; PLCG2; APOA2; BIN1; RAB12; CDH13;\\ APOE; SORL1; TREM2; CLU} \\[10pt]
        Endomembrane system organization & GO:0010256 & $0.000500$ & \makecell[l]{BLOC1S3; NSF; USP8; BIN1; CR1; USP6NL; MTSS2; NECTIN2;\\ CLU} \\[10pt]
        Nitrogen compound transport & GO:0071705 & $0.000741$ & \makecell[l]{BLOC1S3; TOMM40; RABEP1; MME; NUP88; NSF; NUP43; CR1;\\ RAB12; APOE; SORL1; CLU} \\[10pt]
        Organelle localization & GO:0051640 & $0.000771$ & \makecell[l]{BLOC1S3; PICALM; NUP88; TMEM106B; CLU; BIN1; RAB12;\\ NECTIN2} \\[10pt]
        Cellular component organization or biogenesis & GO:0071840 & $0.000792$ & \makecell[l]{BLOC1S3; KLC3; TOMM40; PICALM; RABEP1; MME; NUP88;\\ PTK2B; ALDH1A2; PHB1; TNXB; TMEM106B; NSF; MARK4;\\ KCTD1; CCDC6; CLU; MINDY2; LTBP4; APOA2; USP8; BIN1;\\ CR1; ERCC2; USP6NL; RAB12; GEMIN7; CDH13; APOE; ETV1;\\ MTSS2; NECTIN2; TREM2; CLU} \\[30pt]
        Localization & GO:0051179 & $0.000950$ & \makecell[l]{BLOC1S3; KLC3; TOMM40; PICALM; RABEP1; MME; NUP88;\\ TMEM106B; NSF; FCER1G; SLC24A4; CLU; PLCG2; LTBP4;\\ APOA2; NUP43; BIN1; CR1; USP6NL; RAB12; NBEAL1; CDH13;\\ APOE; SORL1; NECTIN2; TREM2; CLU} \\[5pt]
        \bottomrule 
    \end{tabular}
    \end{adjustbox}
    \label{tab:AD_pos_BO}
\end{table*}
\begin{table*}[!ht]
    \centering
    \caption{Most significant biological processes for MCIc patients obtained from the analysis of the most relevant SNPs with positive IG attributions. In this Table are shown the biological processes, their raw \textit{p}-value, and the overlap genes.}
    \begin{adjustbox}{width=\textwidth}
    \begin{tabular}{ l  l  l  l }
        \toprule
        \makecell[c]{Biological processes} & \makecell[c]{GO term} & \makecell[c]{p-value} & \makecell[c]{Overlap genes}\\
        \midrule
        Positive regulation of amide metabolic process & GO:0034250 & $0.00109$ & \makecell[l]{PICALM; RPS27L; PTK2B; APOE; CLU} \\[5pt]
        Protein-lipid complex assembly & GO:0065005 & $0.00109$ & \makecell[l]{APOC1; ABCA7; APOA2; BIN1; APOE} \\[5pt]
        Regulation of amide metabolic process & GO:0034248 & $0.00128$ & \makecell[l]{PICALM; RPS27L; PTK2B; ABCA7; BIN1; APOE; CELF1;\\ CLU} \\[10pt]
        Positive regulation of endocytosis & GO:0045807 & $0.00147$ & \makecell[l]{FCER1G; ABCA7; PLCG2; APOA2; BIN1; APOE; CLU} \\[5pt]
        Plasma lipoprotein particle assembly & GO:0034377 & $0.00436$ & \makecell[l]{APOC1; ABCA7; APOA2; APOE} \\[5pt]
        Regulation of endocytosis & GO:0030100 & $0.00453$ & \makecell[l]{PICALM; APOC1; FCER1G; ABCA7; PLCG2; APOA2; BIN1;\\ APOE; CLU} \\[10pt]
        Sterol transport & GO:0015918 & $0.00519$ & \makecell[l]{APOC1; ABCA7; APOA2; APOE; CLU} \\[5pt]
        Organic hydroxy compound transport & GO:0015850 & $0.00519$ & \makecell[l]{APOC1; ABCA7; APOA2; APOE; CLU} \\[5pt]
        Cholesterol transport & GO:0030301 & $0.00519$ & \makecell[l]{APOC1; ABCA7; APOA2; APOE; CLU} \\[5pt]
        Protein-lipid complex organization & GO:0071825 & $0.00519$ & \makecell[l]{APOC1; ABCA7; APOA2; BIN1; APOE} \\[5pt]
        Regulation of lipid metabolic process & GO:0019216 & $0.00841$ & \makecell[l]{LACTB; APOC1; ABCA7; PLCG2; APOA2; EPHX2; APOE} \\[5pt]
        Negative regulation of amide metabolic process & GO:0034249 & $0.01440$ & \makecell[l]{ABCA7; BIN1; APOE; CELF1; CLU} \\[5pt]
        Negative regulation of amyloid precursor protein catabolic process & GO:1902992 & $0.01440$ & \makecell[l]{PICALM; ABCA7; BIN1; APOE; CLU} \\[5pt]
        Monoatomic cation homeostasis & GO:0055080 & $0.01440$ & \makecell[l]{PICALM; TMEM106B; SLC24A4; TSPOAP1; APOE} \\[5pt]
        Intracellular monoatomic cation homeostasis & GO:0030003 & $0.01440$ & \makecell[l]{PICALM; TMEM106B; SLC24A4; TSPOAP1; APOE} \\[5pt]
        \bottomrule 
    \end{tabular}
    \end{adjustbox}
    \label{tab:MCIc_pos_BO}
\end{table*}
\begin{table*}[!ht]
    \centering
    \caption{Most significant biological processes for MCInc patients obtained from the analysis of the most relevant SNPs with positive IG attributions. In this Table are shown the biological processes, their raw \textit{p}-value, and the overlap genes.}
    \begin{adjustbox}{width=\textwidth}
    \begin{tabular}{ l  l  l  l }
        \toprule
        \makecell[c]{Biological processes} & \makecell[c]{GO term} & \makecell[c]{p-value} & \makecell[c]{Overlap genes}\\
        \midrule
        T cell activation & GO:0042110 & $0.00695$ & \makecell[l]{FCER1G; TREML2; SPI1; RELB} \\[5pt]
        Lymphocyte activation involved in immune response & GO:0002285 & $0.00695$ & \makecell[l]{FCER1G; PLCG2; ERCC1; RELB} \\[5pt]
        Negative regulation of hydrolase activity & GO:0051346 & $0.00695$ & \makecell[l]{PICALM; APOA2; BIN1; CR1} \\[5pt]
        T cell differentiation & GO:0030217 & $0.00836$ & \makecell[l]{FCER1G; SPI1; RELB} \\[5pt]
        Negative regulation of peptidase activity & GO:0010466 & $0.00836$ & \makecell[l]{PICALM; BIN1; CR1} \\[5pt]
        Negative regulation of endopeptidase activity & GO:0010951 & $0.00836$ & \makecell[l]{PICALM; BIN1; CR1 } \\[5pt]
        Adaptive immune response based on somatic recombination of immune\\ receptors built from immunoglobulin superfamily domains & GO:0002460 & $0.01170$ & \makecell[l]{FCER1G; CR1L; HLA-DRB1; ERCC1; CR1; RELB} \\[5pt]
        Immune effector process & GO:0002252 & $0.01800$ & \makecell[l]{FCER1G; CR1L; PLCG2; ACE; ERCC1; CR1; RELB} \\[5pt]
        Leukocyte mediated immunity & GO:0002443 & $0.02020$ & \makecell[l]{FCER1G; CR1L; ACE; ERCC1; CR1} \\[5pt]
        Male gonad development & GO:0008584 & $0.02850$ & \makecell[l]{ACE; ERCC1; PLEKHA1} \\[5pt]
        Male sex differentiation & GO:0046661 & $0.02850$ & \makecell[l]{ACE; ERCC1; PLEKHA1} \\[5pt]
        Development of primary male sexual characteristics & GO:0046546 & $0.02850$ & \makecell[l]{ACE; ERCC1; PLEKHA1} \\[5pt]
        Antigen processing and presentation of peptide antigen & GO:0048002 & $0.02850$ & \makecell[l]{FCER1G; ACE; HLA-DRB1} \\[5pt]
        Lymphocyte activation & GO:0046649 & $0.03080$ & \makecell[l]{FCER1G; TREML2; PLCG2; ERCC1; SPI1; RELB} \\[5pt]
        Positive regulation of lymphocyte activation & GO:0051251 & $0.03400$ & \makecell[l]{BLOC1S3; HLA-DRB1; SMARCD3; CR1; SPI1} \\[5pt]
        \bottomrule 
    \end{tabular}
    \end{adjustbox}
    \label{tab:MCInc_pos_BO}
\end{table*}

\begin{table*}[!ht]
    \centering
    \caption{Performance comparison of our model with other state-of-the-art methods dealing with missing modalities for AD detection and MCI conversion tasks.}           
    \begin{subtable}{\linewidth}
    \caption{AD detection task. Accuracy (ACC), precision (PRE), and recall (REC) metrics are reported on the test set or averaged during the cross-validation phase ($mean\pm std$).}
    \begin{adjustbox}{width=\linewidth}
    \begin{tabular}{ l  l  l  l  c  l  ccc l}
        \toprule
        \makecell[c]{Authors} & \makecell[c]{Modalities} & \makecell[c]{Study cohort} & \makecell[c]{Input data} &\makecell[c]{Missing data \\ rate} & \makecell[c]{Missing data\\ management} & \makecell[c]{ACC} & \makecell[c]{REC} & \makecell[c]{PRE} & \makecell[c]{XAI} \\
        \midrule
        Venugopalan et al. \cite{venugopalan2021multimodal} & \makecell{sMRI, Genetics,\\
        Clinics} & \makecell{1406 PAT$^{**}$\\ 598 CN} & \makecell{Full MRI volume\\WGS data, Set \\of clinical scores} & \makecell{$\sim67\%$ missing sMRI \\ $\sim61\%$ missing SNP} & \makecell{Zero filling} & \makecell{$0.630^*$\\$0.780^\dagger$} & \makecell{$0.570^*$\\$0.780^\dagger$} & \makecell{$0.620^*$ \\ $0.770^\dagger$}& Yes\\[5pt]
        Liu et al. \cite{liu2021incomplete} & sMRI, FDG-PET & \makecell{85 AD\\ 90 CN} & ROI based features & $50\%$ missing PET & \makecell{Auto-Encoder \\ for complementing \\ complete and incomplete\\ latent representation} & $0.836\pm0.08$ & n.d. & n.d. & No\\[25pt]
        Gao et al. \cite{gao2021task} & sMRI, FDG-PET & \makecell{352 AD\\ 427 CN} & \makecell{sMRI and PET \\ volumes} &  $30\%$ missing PET & \makecell{Task-induced pyramid \\ and attention GAN} & $0.927$ & $0.917$ & n.d.& No\\[10pt]
        Gao et al. \cite{gao2023multimodal} & \makecell{sMRI, FDG-PET,\\T2-MRI} & \makecell{352 AD\\ 427 CN} & \makecell{sMRI, PET,\\T2-MRI volumes} &  \makecell{n.d. $\%$\\ missing data} & \makecell{Multi-level guided GAN} & $0.944$ & $0.930$ & n.d.& Yes\\[10pt]
        Zhang et al. \cite{zhang2022bpgan} & \makecell{sMRI, FDG-PET} & \makecell{345 AD\\ 531 CN} & \makecell{sMRI and PET\\volumes} & \makecell{Generated\\synthetic PET} & \makecell{GAN-based multiple\\convolution U-Net} & $0.981$ & n.d. & n.d. & No \\[10pt]
        Ye et al. \cite{ye2022pairwise} & sMRI, FDG-PET & \makecell{160 AD\\ 210 CN} & \makecell{ROI based features \\ for VBM, FDG} & 50\% missing PET & \makecell{GAN with attention \\ layer in latent space} & $0.914\pm0.19$ & $0.881\pm0.11$ & n.d. & Yes\\[10pt]
        Tu et al. \cite{tu2024multimodal} & sMRI, FDG-PET & \makecell{255 AD\\ 124 CN} & \makecell{sMRI and PET\\volumes} & \makecell{Generated\\synthetic PET} & \makecell{Consistent manifold\\ projection GAN} & $0.980\pm0.003$ & $0.932\pm0.012$ & n.d. & No\\[10pt]
        Zhang et al. \cite{zhang2024pyramid} & sMRI, FDG-PET & \makecell{n.d. AD\\ n.d. CN} & \makecell{sMRI and PET\\volumes} & \makecell{Generated\\synthetic PET} & \makecell{GAN pyramidal\\ attention mechanism} & $0.934$ & $0.975$ & n.d. & Yes\\[10pt]
        Our & \makecell[c]{sMRI, fMRI,\\Genetics} & \makecell{$332$ AD\\ $664$ CN} & \makecell{GM volume \\ sFNC matrix, SNPs} & \makecell{$\sim59\%$ missing fMRI\\ $\sim57\%$ missing SNPs} & \makecell{Multiple cGANs \\ to generate missing \\ latent representation}  & $0.926\pm0.02$ & $0.876\pm0.03$ & $0.910\pm0.05$ & Yes \\%[15pt]
        \bottomrule
    \end{tabular}    
    \end{adjustbox}
    \label{table:classification_CN_AD}
    \end{subtable}
    \begin{subtable}{\linewidth}
    \caption{MCI conversion prediction. Accuracy (ACC), precision (PRE), and recall (REC) metrics are reported on the test set.}
    \begin{adjustbox}{width=\linewidth}
    \begin{tabular}{ l  c  l  c  c  l c ccc l}
        \toprule
        \makecell[c]{Authors} & \makecell[c]{Modalities} & \makecell[c]{Study cohort} & \makecell[c]{Input data} &\makecell[c]{Missing data \\ rate} & \makecell[c]{Missing data\\ management} & \makecell{Trained on the \\ specific task} & \makecell[c]{ACC} & \makecell[c]{REC} & \makecell[c]{PRE} & \makecell[c]{XAI} \\
        \midrule
        Ritter et al. \cite{ritter2015multimodal} & \makecell{$10$ modalities \\ (imaging and clinics)} & \makecell{86 MCIc\\237 MCInc} & Tabular features &$\sim8\%$ missing features & \makecell{Mean imputation and\\imputation by the EM} & Yes & $0.670$ & n.d. & n.d. & No\\[10pt]
        Cai et al. \cite{cai2018deep} & sMRI, FDG-PET & \makecell{76 MCIc\\128 MCInc} & \makecell{GM volume\\PET volumes} & $50\%$ missing PET & \makecell{3D encoder-decoder\\network to generate\\the full PET volume} & Yes & $0.657$ & n.d. & n.d. & No\\[15pt]
        Zhou et al. \cite{zhou2019latent} & \makecell{sMRI, FDG-PET\\Genetics} & \makecell{157 MCIc\\ 205 MCInc} & \makecell{ROI based features\\for PET and MRI,\\ SNPs} & $51\%$ missing PET & \makecell{Latent representation\\learning without\\missing data imputation} & Yes & $0.743$ & n.d. & n.d. & Yes \\[15pt]
        Gao et al. \cite{gao2021task} & sMRI, FDG-PET & \makecell{234 MCIc\\342 MCInc} & \makecell{sMRI and PET \\ volumes} & $30\%$ missing PET & \makecell{Task-induced pyramid \\ and attention GAN} & Yes & $0.753$ & $0.708$ & n.d. & No\\[10pt]
        Gao et al. \cite{gao2023multimodal} & \makecell{sMRI, FDG-PET,\\T2-MRI} & \makecell{234 MCIc\\ 342 MCInc} & \makecell{sMRI, PET,\\T2-MRI volumes} &  \makecell{n.d. $\%$\\ missing data} & \makecell{Multi-level guided GAN} & \makecell{Yes, pre-training\\ on CN vs AD} & $0.778$ & $0.754$ & n.d.& No\\ [10pt]
        Tu et al. \cite{tu2024multimodal} & sMRI, FDG-PET & \makecell{121 MCIc\\ 321 MCInc} & \makecell{sMRI and PET\\volumes} & \makecell{Generated\\synthetic PET} & \makecell{Consistent manifold\\ projection GAN} & Yes & $0.923\pm0.010$ & $0.957\pm0.014$ & n.d. & No\\[10pt]
        Our & \makecell{sMRI, fMRI,\\Genetics} & \makecell{289 MCIc\\646 MCInc} &\makecell{GM volume \\ sFNC matrix, SNPs} & \makecell{$\sim76\%$ missing fMRI\\ $\sim45\%$ missing SNPs} & \makecell{Multiple cGANs \\ to generate missing \\ latent representation} & No & $0.711\pm0.01$ & $0.610\pm0.03$ & $0.558\pm0.03$ & Yes \\
        \bottomrule
    \end{tabular}    
    \end{adjustbox}
    \footnotesize{$^{**}$PAT = 266 AD and 699 MCI, $^*$ CN/PAT (AD+MCI) classification relying on sMRI and SNPs, $^\dagger$ Three classes classification, CN/MCI/AD relying on the three modalities, n.d. = not declared}\\
    \label{table:classification_MCI}
    \end{subtable}%
\end{table*}

\clearpage
\bibliographystyle{unsrt}
\bibliography{arxiv_JNE_GD.bib}